\documentclass[aps,prb,twocolumn,showpacs,amsmath,amssymb]{revtex4}
\usepackage{graphicx}
\usepackage{latexsym}
\usepackage{amsmath}
\usepackage{txfonts}
\usepackage{color}
\usepackage{graphicx}
\usepackage{dcolumn}
\usepackage{bm}
\usepackage{txfonts}
\usepackage{mathrsfs}
\usepackage{feynmf}
\usepackage{comment}
\usepackage{float}

   \let\e=\varepsilon

\def\i{\textrm{i}}
\def\e{\textrm{e}}

\def\cos{\textrm{cos}}

\newcommand{\beq}{\begin{equation}}
\newcommand{\eeq}{\end{equation}}
\newcommand{\ba}{\begin{array}{ccc}}
\newcommand{\ea}{\end{array}}
\newcommand{\nn}{\nonumber \\}

\newcommand{\br}{{\bm r}}
\newcommand{\bs}{{\bm s}}
\newcommand{\bt}{{\bm t}}
\newcommand{\bk}{{\bm k}}
\newcommand{\bp}{{\bm p}}

\def\bea{\begin{eqnarray}}
\def\eea{\end{eqnarray}}

\begin{document}

\title{Deconfined criticality in bilayer graphene}

\author{Junhyun Lee}
\affiliation{Department of Physics, Harvard University, Cambridge MA 02138, USA}

\author{Subir Sachdev}
\affiliation{Department of Physics, Harvard University, Cambridge MA 02138, USA}
\affiliation{Perimeter Institute for Theoretical Physics, Waterloo, Ontario N2L 2Y5, Canada}

\date{\today}

\begin{abstract}
We propose that bilayer graphene can provide an experimental realization of deconfined criticality.
Current experiments indicate the presence of N\'eel order in the presence of a moderate magnetic field.
The N\'eel order can be destabilized by application of a transverse electric field. The resulting electric field induced state is likely to have valence bond solid
order, and the transition can acquire the emergent fractionalized and gauge excitations of deconfined criticality.
\end{abstract}


\maketitle


\section{Introduction}
\label{sec:intro}

Undoped graphene, in both its monolayer and bilayer forms, is nominally a semi-metal.
However, upon application of a moderate magnetic field it turns into an insulator \cite{ong}
(in the quantum Hall terminology, this state has filling fraction $\nu=0$).
Evidence has been accumulating from recent experiments \cite{weitz,freitag,macdonaldexp,young1,maher,basel,young2}
that the insulator has symmetry breaking due to the appearance of antiferromagnetic long-range order.
Because of the applied magnetic field, the antiferromagnetic order is expected to lie in the plane
orthogonal to the magnetic field, along with a ferromagnetic `canting' of the spins along the direction
of the magnetic field: this state is therefore referred to as a canted antiferromagnet (CAF).
For the case of bilayer graphene, experiments \cite{weitz,freitag,macdonaldexp,maher,basel} have also induced what appears to be a quantum phase
transition out of the CAF state. This is done by applying an electric field transverse to the layers, leading to 
states with layer polarization of electric charge, but presumably without antiferromagnetic order. 

\begin{figure}[t]
\includegraphics*[width=27.5mm,angle=0]{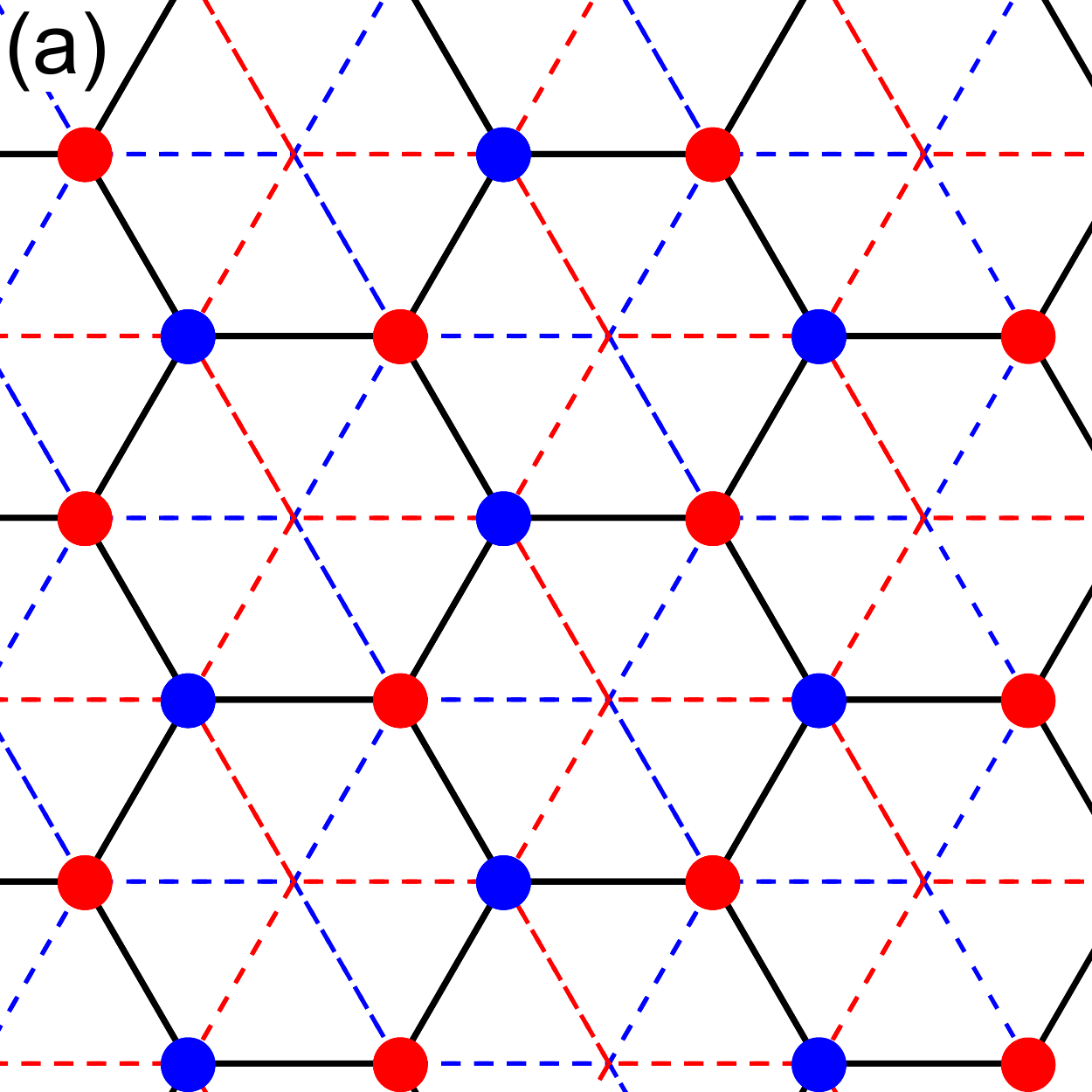}
\hspace{1pt}
\includegraphics*[width=27.5mm,angle=0]{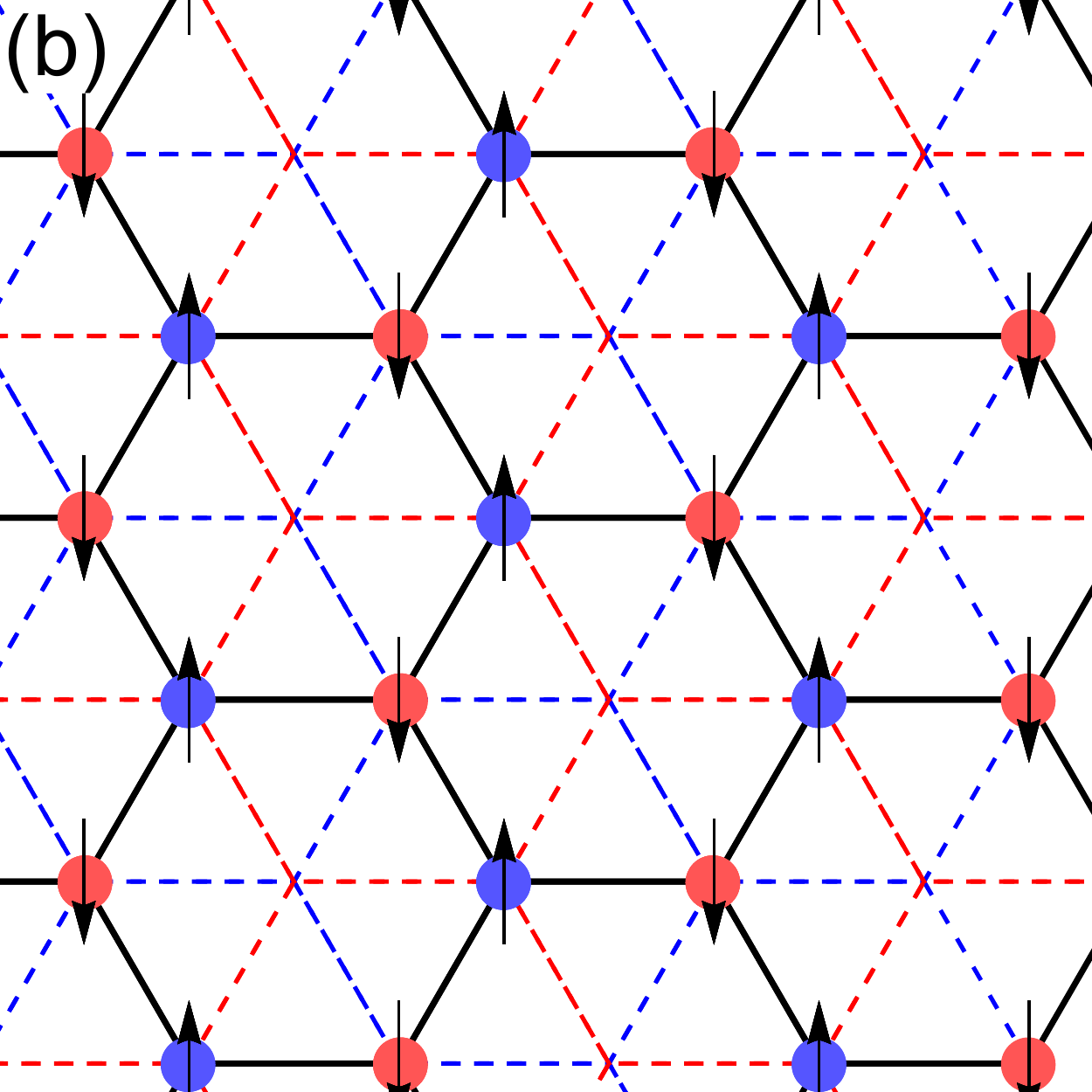}
\hspace{1pt}
\includegraphics*[width=27.5mm,angle=0]{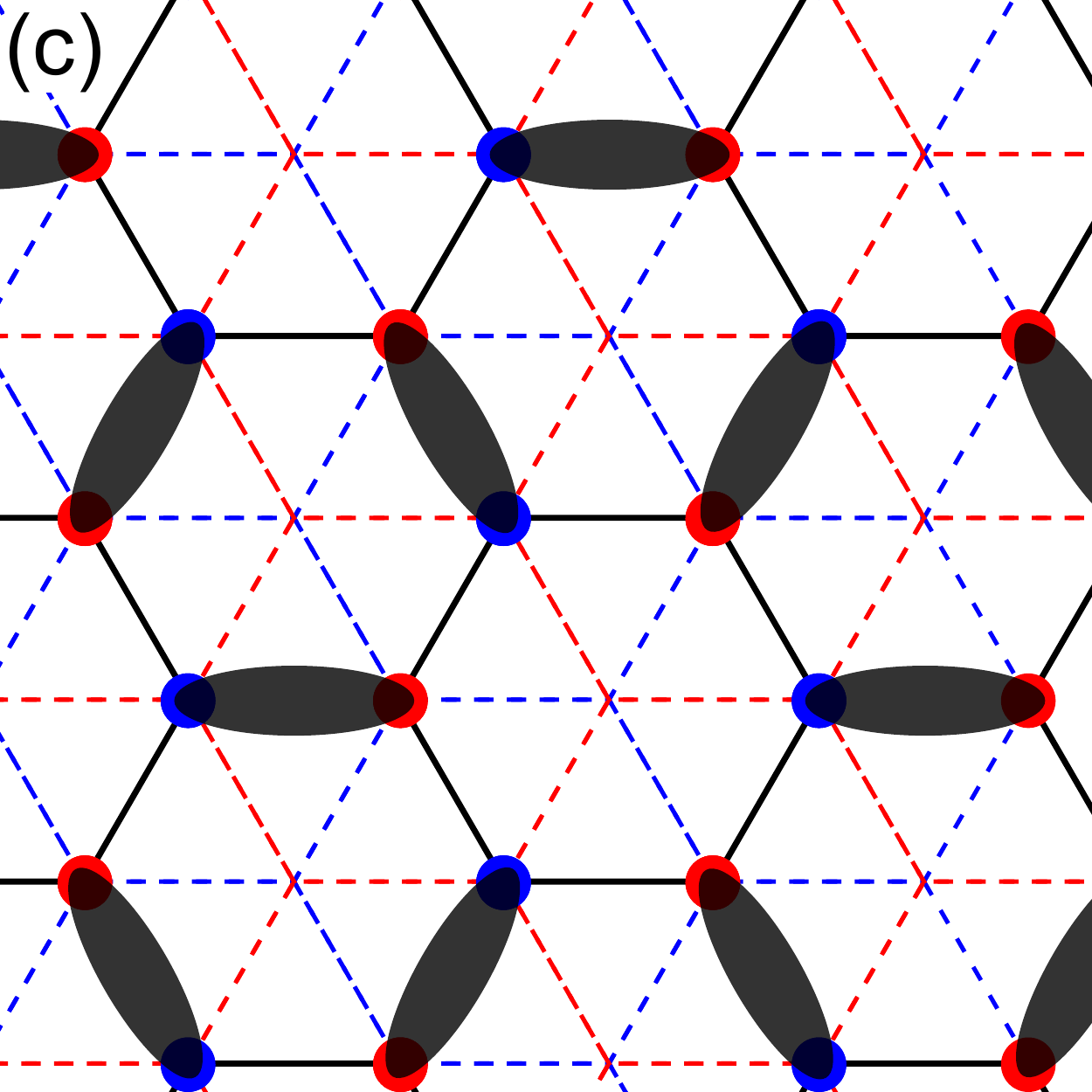}
\caption{(Color online) (a) Top view of $AB$ stacked bilayer graphene. $\Lambda_a$ and $\Lambda_b$ are the sublattice of layer $A$ of graphene which is the dashed red line in the figure; the sites of $\Lambda_b$ are colored blue. And $\Lambda_c$ and $\Lambda_d$ are the sublattice of layer $B$ which is the dashed blue line; the sites of $\Lambda_c$ are colored red. The sites of  $\Lambda_a$ and $\Lambda_d$ are present at the same $\br$ in the plane of graphene. $\Lambda_b$ and $\Lambda_c$ also makes a honeycomb lattice, which is depicted as the thick black line. This is the effective honeycomb lattice where the N\'{e}el and VBS order reside. (b) 
N\'{e}el order in the effective honeycomb lattice. (c) One of the three VBS states in the effective lattice. The black oval depicts the singlet bonds.
Note that 1/3 of the hexagons in the VBS state have no valence bonds, and so this state can also be viewed as having `plaquette' order on these hexagons.}
\label{fig:blg}
\end{figure}

Theoretically, the CAF is expected to be stable in bilayer graphene over a range of microscopic parameters \cite{vafek,allan,levitov}.
Studying the instability of the CAF in a Hartree-Fock analysis, Kharitonov \cite{khari1,khariprl,khari2,khari3} 
proposed phase diagrams which apply to the experimental
configurations: he found that upon application of an electric field, the CAF state undergoes
a quantum phase transition into a state with partial-layer-polarization (PLP) and a distinct broken symmetry:
the PLP state preserves spin rotation invariance, but breaks lattice symmetries in the 
Kekul\'{e} pattern (see Fig.~\ref{fig:blg}c). 
The insulating CAF and PLP states, and the transition between them, will be the focus of our present study.
There is no direct experimental evidence yet for the 
Kekul\'{e} broken symmetry in the PLP state, but we hope this will be the focus of future experiments.

From the perspective of symmetry, we are therefore investigating the quantum phase transition between
two insulating states in an electronic model which has spin rotation symmetry and the  
space group symmetries of the honeycomb lattice. One insulator breaks spin rotation symmetry by the
appearance of antiferromagnetic long-range order in the two-sublattice pattern shown in Fig.~\ref{fig:blg}b:
we will henceforth refer to this insulator as the N\'eel state. The second insulator breaks the space group
symmetry alone in the Kekul\'e pattern of Fig.~\ref{fig:blg}c. A direct quantum phase transition between
two insulators with precisely the same symmetries was first discussed some time ago in Ref.~\onlinecite{vbsprb}
in the very different context of correlated electron models inspired by the cuprate high temperature superconductors.
In these models, the Kekul\'e state is referred to as a valence bond solid (VBS), as the space group symmetry is broken
by singlet valence bonds between spins on the sites of the honeycomb lattice; we include the `plaquette' resonating
state within the class of VBS states, and it breaks the honeycomb lattice symmetry in the same pattern.

More recently, the N\'eel-VBS transition has been identified \cite{senthil1,senthil2} as a likely candidate for `deconfined criticality'.
In this theory, the low energy excitations in the vicinity of the transition are described by 
neutral excitations carrying spin $S=1/2$ (`spinons') coupled to each other by the `photon' of an emergent
U(1) gauge field. The quantum transition itself is either second order or weakly first order; in either case,
there is evidence for the presence of the emergent gauge excitations \cite{sandvik,ribhu}.

In the present paper, we will apply a strong coupling perspective to models on the bilayer honeycomb
lattice linked to the physics of bilayer graphene. Our analysis therefore complements that of Kharitonov, who
perturbatively examined the effect of interactions after projecting to the lowest Landau level.
We also note other theoretical studies by Roy and collaborators \cite{roy1,roy2,roy3} which do not project to the lowest Landau level.
Our perspective is more suited to addressing the nature of quantum fluctuations near the 
quantum phase transition, and for describing the possible emergence of exotic varieties of fractionalization.
We will discuss some of the experimental consequences of this new perspective in Section~\ref{sec:conc}. 

We will introduce our lattice model on the bilayer honeycomb lattice in Section~\ref{sec:model}. 
We assume that the strongest coupling in the model is the on-site Hubbard repulsion $U$, and perform a traditional
$1/U$ expansion to obtain an effective spin model on the same lattice. 
In Section~\ref{sec:spinwave}, we examine this spin model in a spin-wave expansion, and determine regimes where
the N\'eel order is suppressed. An alternative effective spin model, related to those examined in recent numerical work,
is studied in Section~\ref{sec:j1j2}. We study the geometric phases between the N\'eel and VBS orders in Section~\ref{sec:topo},
and comment on the structure of vortices in the VBS order in Section~\ref{sec:vortex}.

\section{The strong coupling model}
\label{sec:model}

We start our analysis from the extended Hubbard model for $AB$ stacked bilayer graphene in the strong coupling limit. A top view of $AB$ stacked bilayer graphene is shown in Fig.~\ref{fig:blg}a. In our coordinate system, we set the lattice constant to $1$ and define $\bs_1 = (1,0)$, $\bs_2 = (-1/2,\sqrt{3}/2)$,
and $\bs_3 = (-1/2, -\sqrt{3}/2  )$. 
\begin{align}
	H =& -t_{\parallel} \sum_{\br \in \Lambda_a} \sum_{i=1}^3 c^{(a)\dagger}_{\br}  c^{(b)}_{\br+\bs_i}  
	-t_{\parallel} \sum_{\br \in \Lambda_d} \sum_{i=1}^3 c^{(d)\dagger}_{\br}   c^{(c)}_{\br-\bs_i}   
	-t_{\perp} \sum_{\br \in \Lambda_a} c^{(a) \dagger}_{\br} c^{(d)}_{\br}
	\nonumber \\
	&+\textrm{h.c.}   +U \sum_{\alpha=a, b, c, d}  \sum_{\br \in \Lambda_\alpha} \frac{n^{(\alpha)}_{\br} (n^{(\alpha)}_{\br} -1 )}{2} \nonumber \\
	& + V_{\parallel}  \sum_{\br \in \Lambda_a} \sum_{i=1}^3 n^{(a)}_{\br} n^{(b)}_{\br + \bs_i}
	+ V_{\parallel}  \sum_{\br \in \Lambda_d} \sum_{i=1}^3  n^{(d)}_{\br} n^{(c)}_{\br - \bs_i} 
	+ V_{\perp} \sum_{\br \in \Lambda_a} n^{(a)}_{\br} n^{(d)}_{\br}  \nonumber \\
	&+ E \left( - \sum_{\br \in \Lambda_a} n^{(a)}_{\br} - \sum_{\br \in \Lambda_b} n^{(b)}_{\br} + \sum_{\br \in \Lambda_c} n^{(c)}_{\br} + \sum_{\br \in \Lambda_d} n^{(d)}_{\br}\right)
	\label{eq:hubbard_H}
\end{align}
Here $c^{\dagger}$ ($c$) is the fermion creation (annihilation) operator and $n = c^{\dagger} c$ is the number operator. $t_\parallel$ and $V_\parallel$ are the tight binding hopping parameter and the nearest neighbor interaction within the plane, $t_\perp$ and $V_\perp$ are those between the planes, and $U$ is the on-site interaction. We label each layer of graphene as $A$ and $B$: layer $A$ consists of sublattice $\Lambda_a$ and $\Lambda_b$, and layer $B$ consists of sublattice $\Lambda_c$ and $\Lambda_d$. Only one of the sublattice in each layer has common in-plane coordinate in $AB$ stacked bilayer graphene, and we set those to be $\Lambda_a$ and $\Lambda_d$.
Elsewhere in the literature, the site labels $a$, $b$, $c$, and $d$ are often referred to as $A1$, $B1$, $A2$, and $B2$ respectively, meaning sublattice $A$($B$) or layer 1(2). However, we find it more convenient to use the compact notation $a$, $b$, $c$, $d$. Hopping and interaction between the layers only occur between these sublattices. We also include an electric field transverse to the plane of graphene, pointing from layer $A$ to layer $B$. The electric field is minimally coupled to the density of the fermions with coupling $E$.  We assume that $E$ is also
smaller than $U$, and so both layers will be half-filled at leading order in $1/U$, and the effective Hamiltonian can be expressed only in terms
of spin operators on the sites. The subleading $1/U$ corrections will induce terms in the effective Hamiltonian, but also induce a polarization in the layer
density when computed in terms of the bare electron operators. 

Our Hamiltonian does not explicitly include the influence of an applied magnetic field. Such a field will modify $H$ in two ways, via a Peierls phase factor on the hopping terms $t_{\parallel,\perp}$, and a Zeeman coupling. In the context of our strong coupling expansion, the influence of the Peierls phases will only be to modify the coefficients of ring-exchange terms in the effective spin Hamiltonian. However, such ring-exchange terms only appear at sixth order in $t_\parallel$, and this is
higher order than our present analysis; so we can safely drop the Peierls phases. The Zeeman term commutes with all other terms in $H$, and so does not
modify the analysis below, and can be included as needed in the final effective Hamiltonian.

From this Hamiltonian we work on the strong coupling limit, where $t_\parallel, t_\perp \ll U, V$, and perform the $t/U$ expansion up to $\mathcal O (t^4 /U^3 )$ order. In this expansion we assume both $t_\parallel$ and $t_\perp$ are much smaller than $U$, although this is not well satisfied in the experiment (also,
there is a significant difference in the values of the hopping parameters,\cite{zhang} $t_\parallel \sim 3.0$ eV, $t_\perp \sim 0.40$ eV). There are numerous works on the $t/U$ expansion of Hubbard model in various lattices, including the classic work of Ref.s~\onlinecite{macdonald, takahashi} in square lattice. Extra care is needed while dealing the similar procedure with the above model since we have included nearest neighbor interaction and the lattice structure is more complicated. 

First we organize the Hamiltonian in Eq.~\ref{eq:hubbard_H} as $H = H_U + H_t$, where $H_U$ is the interaction terms and $H_t$ is the kinetic terms. We consider $H_t$ as the perturbation and rearrange it by the change of interaction energy through the hopping process. 
\begin{align}
		H_t = \sum_{\lambda} \left[ T_\lambda + T_{-\lambda} \right]
\end{align}
$T_\lambda$ is the sum of all hopping terms that increases the interaction energy by $\lambda U$. For notational convenience, we restrict $\lambda$ to be positive and collect the decreasing energy processes to $T_{-\lambda}$ with an explicit negative sign. 

By systematically performing the unitary transformation, we may obtain the effective Hamiltonian $H^{(n)}$ which contains terms up to the order of $t^{n+1} / U^n$ for arbitrary $n$ \cite{macdonald, takahashi}. We present the result of $H^{(3)}$ for the system in the ground state manifold at half filling without long derivation.
\begin{align}
	H^{(3)}_{GS, HF} = 
	&-\frac{1}{U} \sum_{\lambda} \frac{1}{\lambda } T_{-\lambda} T_{\lambda} \nonumber \\
	&+\frac{1}{2 U^3} \sum_{\lambda_1 , \lambda_2} \frac{1}{\lambda_1 \lambda_2}\left( \frac{1}{\lambda_1} + \frac{1}{\lambda_2} \right) 
	T_{-\lambda_1} T_{\lambda_1} T_{-\lambda_2} T_{\lambda_2}  \nonumber \\
	&-\frac{1}{U^3} \sum_{\lambda_1 + \lambda_2 = \lambda_3 + \lambda_4 \atop \lambda_1 \neq \lambda_3} \frac{1}{\lambda_1 \lambda_4 (\lambda_1 - \lambda_3)}
	T_{-\lambda_1} T_{\lambda_3} T_{-\lambda_2} T_{\lambda_4}  \nonumber \\
	&-\frac{1}{U^3} \sum_{\lambda_1 + \lambda_2 = \lambda_3 + \lambda_4} \frac{1}{\lambda_1 \lambda_4 (\lambda_1 + \lambda_2)}
	T_{-\lambda_1} T_{-\lambda_2} T_{\lambda_3} T_{\lambda_4} 
	\label{eq:H3}
\end{align}
The above expression is a general result for Hubbard type Hamiltonian and will hold for any bipartite lattice regardless of the dimension. 

\begin{figure}[t]
\includegraphics*[width=57mm,angle=0]{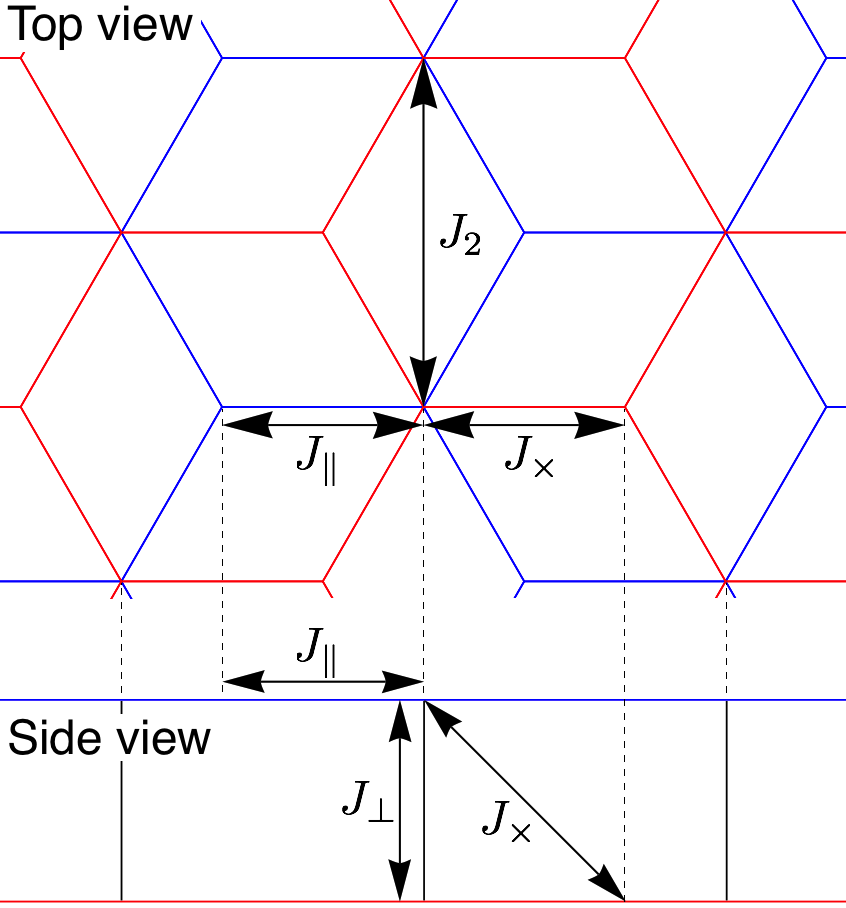}
\caption{(Color online) Top and side view of the $AB$ stacked bilayer graphene. Exchange couplings $J_\parallel$, $J_\perp$, $J_2$, and $J_\times$ are shown in the figure. The top view can be considered as the same as Fig.~\ref{fig:blg}a, without the effective honeycomb lattice depicted in thick black line. Layer $A$ and $B$ are the red and is the blue lattices as in Fig.~\ref{fig:blg}. In the side view, the black lines depict the lattice sites connected by the $t_\perp$ hopping parameter in Eq.~\ref{eq:hubbard_H}. The dashed lines are guide to the eye showing that the horizontal coordinates are the same for the two views. As we can see from the figure, $J_\parallel$ is between nearest neighbors within one layer, $J_\perp$ is between nearest neighbors of different layers, $J_2$ is between next nearest neighbors within one layer, and $J_\times$ is between next nearest neighbors of different layers.}
\label{fig:coupling}
\end{figure}

Applying Eq.~\ref{eq:H3} to our specific case of bilayer graphene we obtain a spin Hamiltonian which contains every terms up to the order of $t^4/U^3$,
\bea
H &=& J_{\parallel} \sum_{\br \in \Lambda_a} \sum_{i=1}^3 \vec{S}^{(a)}_{\br} \cdot  \vec{S}^{(b)}_{\br+\bs_i} + J_{\parallel} \sum_{\br \in \Lambda_d} \sum_{i=1}^3 \vec{S}^{(d)}_{\br} \cdot  \vec{S}^{(c)}_{\br-\bs_i} \nn
 &+& J_{\perp} \sum_{\br \in \Lambda_a}  \vec{S}^{(a)}_{\br} \cdot  \vec{S}^{(d)}_{\br} 
+ J_{2} \sum_{\alpha = a, b, c, d}  \sum_{\br \in \Lambda_\alpha} \sum_{i=1}^3  \vec{S}^{(\alpha)}_{\br} \cdot  \vec{S}^{(\alpha)}_{\br+\bt_i} \nn 
&+&
J_{\times} \sum_{\br \in \Lambda_a} \sum_{i=1}^3 \vec{S}^{(a)}_{\br} \cdot  \vec{S}^{(c)}_{\br-\bs_i} + J_{\times} \sum_{\br \in \Lambda_d} \sum_{i=1}^3 \vec{S}^{(d)}_{\br} \cdot  \vec{S}^{(b)}_{\br+\bs_i},
\label{eq:spinH}
\eea
with the exchange couplings as,
\bea
	J_{\parallel} &=& \frac{4 ~t^2_{\parallel} }{U-V_{\parallel} - \left( \frac{V^2_{\perp}}{U-V_{\parallel}} \right)} , 
	\hspace{5mm}
	J_{\perp} = \frac{4~t^2_{\perp}}{U-V_{\perp} - \left( \frac{E^2}{U-V_{\perp}} \right)}, \nn
	J_2 &=& \frac{4~t^4_{\parallel}}{(U-V_{\parallel})^2 - V^2_{\perp}} \left( \frac{2(U-V_{\parallel}) \left( (U-V_{\parallel})^2 + V^2_{\perp}\right)}{\left( (U-V_{\parallel})^2 - V^2_{\perp} \right)^2} -\frac{1}{U} \right), \nn
	\label{eq:couplings1}
\eea
and, 
\begin{widetext}
\begin{align}
	J_\times =& \frac{4 ~ t^2_\parallel \, t^2_\perp}{\left( (U - V^2_\parallel )^2 - V^2_\perp \right)^2 \left( \left(U-V_\perp \right)^2 - E^2 \right)^2 \left( (U+ V_\perp)^2 - E^2\right)} \times \nn
	&\left( (U-V_\perp )^2 (U+V_\perp ) \left( (U^2 +  V_\perp U - V_\parallel V_\perp)(U - V_\parallel)^2 
	 - V^2_\perp (U^2 - (4 V_\parallel - V_\perp ) U + (2 V_\parallel - V_\perp ) V_\parallel )  \right)  \right.  \nn
	&\left. - E^2 \left( 2U^5 -(4 V_\parallel + 5 V_\perp ) U^4 + (2 V^2_\parallel + 13 V_\parallel V_\perp + 8 V^2_\perp)U^3- V_\perp (V_\parallel + V_\perp )(11 V_\parallel + 2V_\perp ) U^2 \right. \right. \nn
	&\left. \left. +V_\perp (V_\parallel + V_\perp )(3V_\parallel - V_\perp) (V_\parallel + 2 V_\perp) U - V^2_\perp (V^3_\parallel + 3 V^2_\parallel V_\perp - V_\parallel V^2_\perp + V^3_\perp)\right)  + E^4 (U- V_\perp) ((U-V_\parallel)^2+ V^2_\perp) \right)
	\label{eq:couplings2}
\end{align}
\end{widetext}
We have additionally defined $\bt_1 = \bs_2 - \bs_3$,  $\bt_2 = \bs_3 - \bs_1$, and $\bt_3 = \bs_1 - \bs_2$. Without the electric field, the lattice symmetry guarantees the $J_\times$ coupling in $\vec{S}^{(a)}_{\br} \cdot  \vec{S}^{(c)}_{\br-\bs_i}$ and $ \vec{S}^{(d)}_{\br} \cdot  \vec{S}^{(b)}_{\br+\bs_i}$ to be the same. However, when the field is turned on, the layer symmetry breaks and the two $J_\times$ value becomes different. Here we take the average value for simplicity. This will not change the qualitative behavior unless $E$ is very large. Different exchange couplings defined in Eq.~\ref{eq:spinH} are shown schematically in Fig.~\ref{fig:coupling}.
We work in the parameter range where all four exchange couplings are antiferromagnetic. This can be made compatible with experimental data of hopping parameters \cite{zhang}. Moreover, in most of the parameter regime where $J_\parallel$ and $J_\perp$ are antiferromagnetic, we find $J_2$ and $J_\times$ to be positive as well. Therefore, we have enough parameter space to explore with this model and do not have to fine-tune the parameters.  

\section{Spin-wave expansion}
\label{sec:spinwave}

Previous studies of the bilayer antiferromagnet have focused on the square lattice \cite{hida,sandvik1,morr,millis,sandvik2} where the sites are stacked directly on top of each other. In these models, as the interlayer coupling is increased there is eventually a transition from the N\'eel state to a `trivial' paramagnet in which
the ground state is approximately the product of interlayer valence bonds between superposed spins.
However, here we are considering a staggered stacking, in which no such trivial one-to-one identification of spins between the two layers is possible.
Any pairing of spins must break a lattice symmetry, and this is a simple argument for the appearance of a VBS state.
Nevertheless, it is useful to apply the spin wave expansion used for the square lattice \cite{hida,morr}, and study how the intra-
and inter-layer couplings modify the staggered magnetization. This will help us determine the parameter regime over which the 
N\'eel order decreases, and a possible VBS state can appear. However, a description of the transition to, and structure of, the VBS state
is beyond the regime of applicability of the spin-wave expansion.

Among the four exchange couplings listed in Eq.~\ref{eq:couplings1} and \ref{eq:couplings2}, only $J_\perp$ and $J_\times$ depend on the electric field strength, $E$. The electric field breaks the layer symmetry, so it is reasonable that $E$ is only included in the exchange coupling between different layers. We start from the N\'{e}el phase and calculate the staggered magnetization of the bilayer graphene as a function of either $J_\perp$, $J_\times$, or $E$. Since our starting point is an SU(2) symmetry broken state, we use the Holstein-Primakoff representation.

Starting from the effective spin Hamiltonian derived in Eq.~\ref{eq:spinH}, we perform the $1/S$ expansion (where $S$ is the magnitude of the spin, and we are interested in $S=1/2$) 
about the antiferromagnetically ordered state by expressing the spin operators in terms of bosons,
$a$, $b$, $c$, $d$:
%
\begin{alignat}{3}
&S_z^{(a)} = S - a^\dagger a  \quad &;& \quad S_+^{(a)} = \sqrt{2S} (1 - a^\dagger a /(2S))^{1/2} a \nn
&S_z^{(b)} = -S + b^\dagger b \quad &;& \quad S_+^{(b)} = \sqrt{2S} b^\dagger (1 - b^\dagger b /(2S))^{1/2} \nn
&S_z^{(c)} = S - c^\dagger c  \quad &;& \quad S_+^{(c)} = \sqrt{2S} (1 - c^\dagger c /(2S))^{1/2} c \nn
&S_z^{(d)} = -S + d^\dagger d \quad &;& \quad S_+^{(d)} = \sqrt{2S} d^\dagger (1 - d^\dagger d /(2S))^{1/2} \nn
\end{alignat}
%
Then, to the needed order, the Hamiltonian is,
\begin{align}
H &= J_\parallel \sum_{\br \in \Lambda_a} \sum_{i=1}^3 \left( - S^2 + S( a_{\br}^\dagger a_{\br} + b_{\br + \bs_i}^\dagger b_{\br + \bs_i} + a_{\br}^\dagger b_{\br + \bs_i}^\dagger + a_{\br} b_{\br + \bs_i}) \right) \nn
&+ J_\parallel \sum_{\br \in \Lambda_d} \sum_{i=1}^3 \left( - S^2 + S( d_{\br}^\dagger d_{\br} + c_{\br - \bs_i}^\dagger c_{\br - \bs_i} + d_{\br}^\dagger c_{\br - \bs_i}^\dagger + d_{\br} c_{\br - \bs_i}) \right) \nn
&+ J_\perp \sum_{\br \in \Lambda_a}  \left( - S^2 + S( a_{\br}^\dagger a_{\br} + d_{\br}^\dagger d_{\br} 
 + a_{\br}^\dagger d_{\br}^\dagger + a_{\br} d_{\br} \right) \nn
 &+ J_2 \sum_{\br \in \Lambda_a} \sum_{i=1}^3 \left(  S^2 + S( -a_{\br}^\dagger a_{\br} - a_{\br + \bt_i}^\dagger a_{\br + \bt_i} + a_{\br}^\dagger a_{\br + \bt_i} +  a_{\br + \bt_i}^\dagger a_{\br}) \right) \nn
&+ J_2 \sum_{\br \in \Lambda_b} \sum_{i=1}^3 \left(  S^2 + S( -b_{\br}^\dagger b_{\br} - b_{\br + \bt_i}^\dagger b_{\br + \bt_i} + b_{\br}^\dagger b_{\br + \bt_i} + b_{\br + \bt_i}^\dagger b_{\br} ) \right) \nn
&+ J_2 \sum_{\br \in \Lambda_c} \sum_{i=1}^3 \left(  S^2 + S( -c_{\br}^\dagger c_{\br} - c_{\br + \bt_i}^\dagger c_{\br + \bt_i} + c_{\br}^\dagger c_{\br + \bt_i}+  c_{\br + \bt_i}^\dagger c_{\br}) \right) \nn
&+ J_2 \sum_{\br \in \Lambda_d} \sum_{i=1}^3 \left(  S^2 + S( -d_{\br}^\dagger d_{\br} - d_{\br + \bt_i}^\dagger d_{\br + \bt_i} + d_{\br}^\dagger d_{\br + \bt_i} +  d_{\br + \bt_i}^\dagger d_{\br}) \right) \nn
&+ J_\times \sum_{\br \in \Lambda_a} \sum_{i=1}^3 \left(  S^2 + S( - a_{\br}^\dagger a_{\br} - c_{\br - \bs_i}^\dagger c_{\br - \bs_i} + a_{\br}^\dagger c_{\br - \bs_i}+ c_{\br - \bs_i}^\dagger a_{\br} ) \right) \nn
&+ J_\times \sum_{\br \in \Lambda_d} \sum_{i=1}^3 \left(  S^2 + S( - d_{\br}^\dagger d_{\br} - b_{\br + \bs_i}^\dagger b_{\br + \bs_i} + d_{\br}^\dagger b_{\br + \bs_i} + b_{\br + \bs_i}^\dagger d_{\br} ) \right) . \nn
\end{align}
We write this in momentum space as,
\beq
H = -3 N S(S+1) (2 J_\parallel + J_\perp/3 - 4 J_2 - 2 J_\times) + S \sum_\bk \Psi_{\bk}^\dagger M(\bk) \Psi_k ,
\eeq
where $N$ is the number of sites in $\Lambda_a$, $\Psi_{\bk}$ is the boson spinor $\Psi_{\bk} = (a_\bk , c_\bk, b^{\dagger}_{-\bk}, d^\dagger_{-\bk})$, and
\begin{align}
M(\bk) = \left( \begin{array}{cccc}
\tilde{J} (\bk) + J_\perp ~~&~~ J_\times \gamma (-\bk) ~~&~~ J_\parallel \gamma (\bk) ~~&~~  J_\perp \\
J_\times \gamma (\bk) ~~&~~ \tilde{J} (\bk) ~~&~~ 0 ~~&~~ J_\parallel \gamma (\bk) \\
J_\parallel \gamma (-\bk) ~~&~~ 0 ~~&~~ \tilde{J} (\bk) ~~&~~ J_\times \gamma (-\bk) \\
 J_\perp ~~&~~ J_\parallel \gamma (-\bk) ~~&~~ J_\times \gamma (\bk) ~~&~~ \tilde{J} (\bk) +  J_{\perp}
\end{array}\right), \nonumber
\end{align}
with
\bea
\tilde{J} (\bk)  &=& J_\parallel + J_2 \Gamma (\bk) - 3 J_{\times} , \nn
\gamma (\bk) &=& \sum_{i = 1}^3 e^{i \bk \cdot \bs_i } , \nn
\Gamma(\bk) &=& - 6  + 2 \sum_{i=1}^3 \cos \left( \bk \cdot \bt_i \right). \nonumber
\eea
The Hamiltonian can be diagonalized by a bosonic version of the Bogoliubov transformation (which is not a unitary transformation)
as described in Ref.~\onlinecite{sskagome}. 

\begin{figure}[t]
\centering
	\vspace*{2mm}
\includegraphics[width=3.1in]{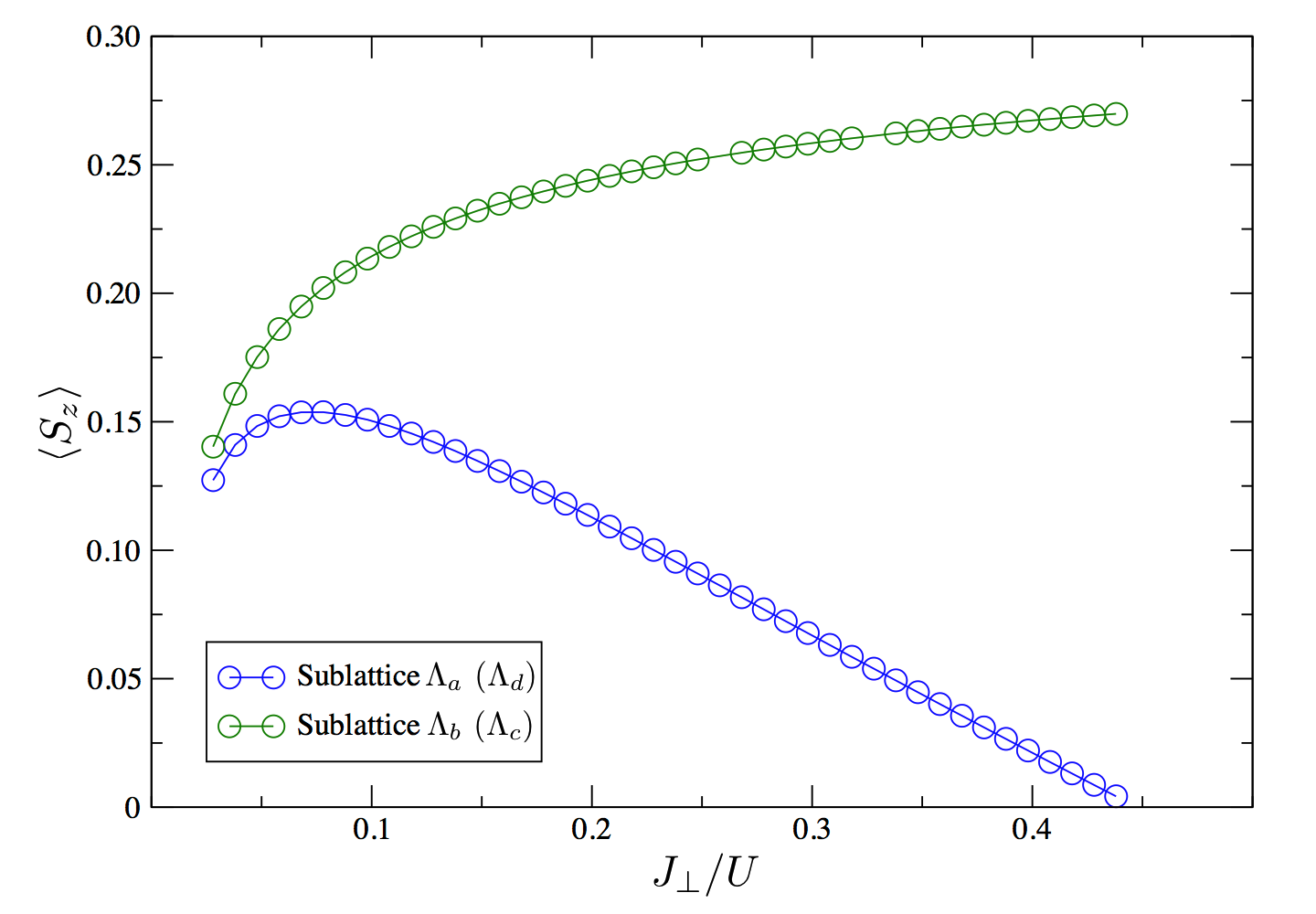}
\includegraphics[width=3.1in]{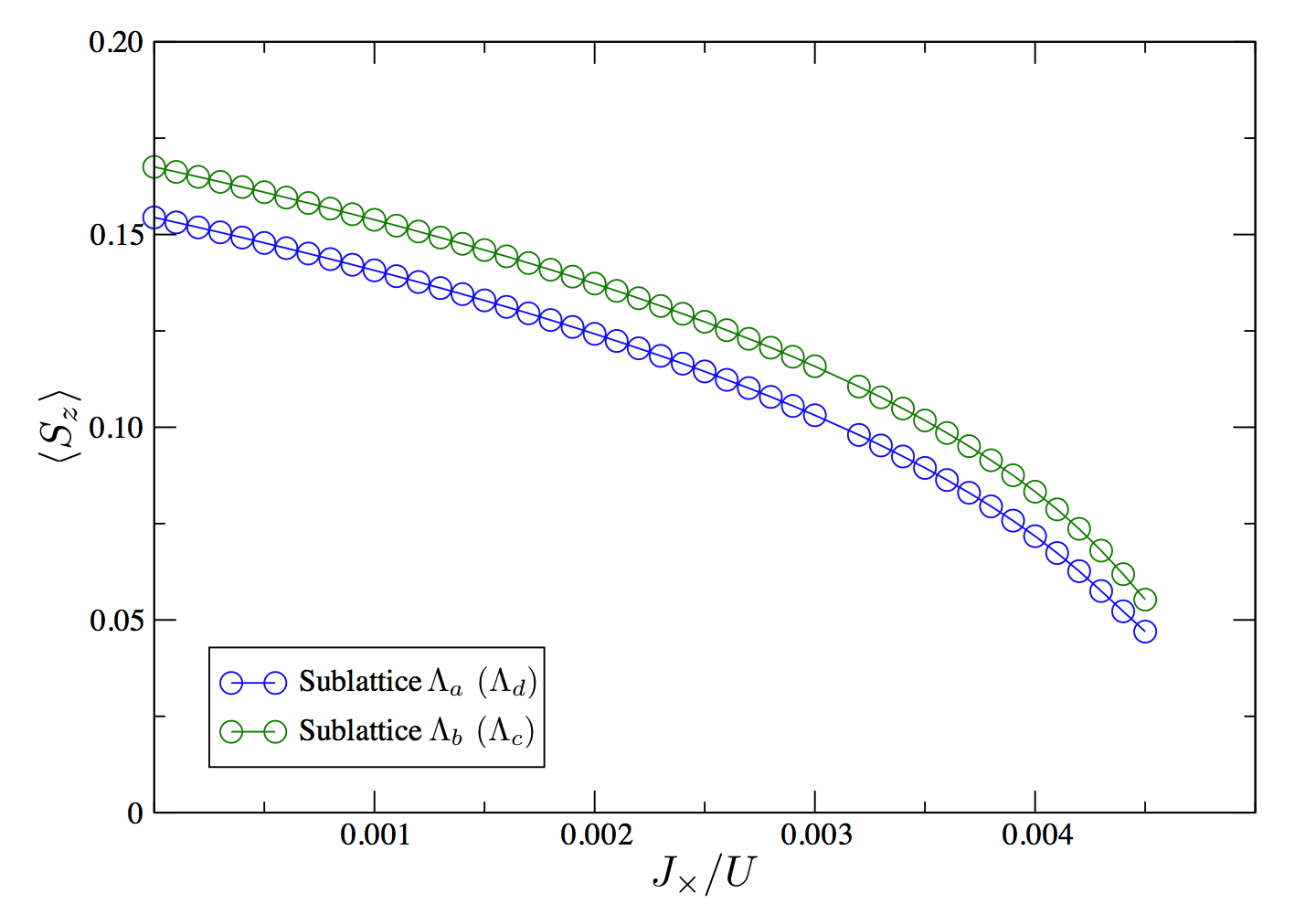}
\caption{(Color online) Magnetization of each sublattice in bilayer graphene as a function of $J_\perp$ (left) and $J_\times$ (right). We used $J_\parallel /U = 0.089$, $J_2 /U = 0.0095$ for both plots, $J_\times /U = 0.0018$ for the left plot, and $J_\perp /U = 0.028$ for the right plot.}
\label{fig:mag-j}
\end{figure}

Now the staggered magnetization of the bilayer graphene can be obtained from the diagonalized Hamiltonian. The expression for the magnetization is very complicated with all four exchange couplings, and hard to write down in a closed form. Therefore we present numerical values for a selected set of parameters. Fig.~\ref{fig:mag-j} shows the calculated magnetization as a function of $J_\perp$ and $J_\times$ for parameters $J_\parallel /U = 0.089$, $J_2 /U = 0.0095$, $J_\times /U = 0.0018$, and $J_\perp /U = 0.028$ (unless one is the variable for the graph). 
These correspond to $t_\parallel / U = 0.1$, $t_\perp /U = 0.07$, $V_\parallel /U = 0.4$, and $V_\perp /U = 0.3$ for the parameters in the extended Hubbard model. Since the sublattice $\Lambda_a$($\Lambda_d$) and $\Lambda_b$($\Lambda_c$) are not symmetric in $AB$-stacked bilayer graphene, they will in general have different magnetization, and therefore are plotted separately (for example, sites in $\Lambda_a$($\Lambda_d$) has coordination number of 4, whereas the sites in $\Lambda_b$($\Lambda_c$) has 3). As depicted in Fig.~\ref{fig:blg}, the N\'{e}el and VBS states of interest reside in $\Lambda_b$ and $\Lambda_c$. We are therefore more interested in the magnetization of $\Lambda_b$ than $\Lambda_a$. 

We observe that the magnetization in each sublattice decreases as $J_\times$ increases. This is reasonable in the sense that antiferromagnetic $J_\times$ increases frustration of the N\'{e}el phase. However when $J_\perp$ increases, magnetization of $\Lambda_b$ increases while that of $\Lambda_a$ decreases. This is mainly because $J_\times$ frustrates all four sublattices, but $J_\perp$ only gives frustration to $\Lambda_a$ and $\Lambda_d$. That is, the N\'{e}el phase in $\Lambda_b$ and $\Lambda_c$ are not directly effected by $J_\perp$. We will explore again, in the next section, 
the influence of $J_\perp$ to $\Lambda_b$ and $\Lambda_c$ in the subleading order in a perturbation theory in large $J_\perp$. The result 
is an antiferromagnetic coupling between spins in $\Lambda_b$ and $\Lambda_c$, and a ferromagnetic coupling between spins within $\Lambda_b$ or $\Lambda_c$. This suggests a `layer-polarized antiferromagnet' state where, for example, every spin in $\Lambda_b$ is polarized up, 
and every spin on $\Lambda_c$ is polarized down. So the increased magnetization of $\Lambda_b$ can be explained in this manner. This is not the scenario we expect in the N\'{e}el to VBS phase transition, because the N\'{e}el order is becoming stronger with an increase of $J_\perp$.  
However, there will not be a case where $J_\perp$ increases alone because $J_\times$ coupling will also increase as we increase the electric field. This $J_\times$ gives frustration to the layer-polarized order which will result in the decrease of the staggered magnetization. 
Note that, in any case, the magnetization is smaller for $\Lambda_a$ than $\Lambda_b$, which contradicts our usual intuition that larger coordination number agrees better with mean field result. However, this result is in accordance with Ref.~\onlinecite{langblg}, where they find the same behavior by quantum Monte Carlo simulation for a Heisenberg model with only $J_\parallel$ and $J_\perp$ couplings, but in a wide range of $J_\perp$. 

\begin{figure}
\centering
	\vspace*{2mm}
\includegraphics[width=3.1in]{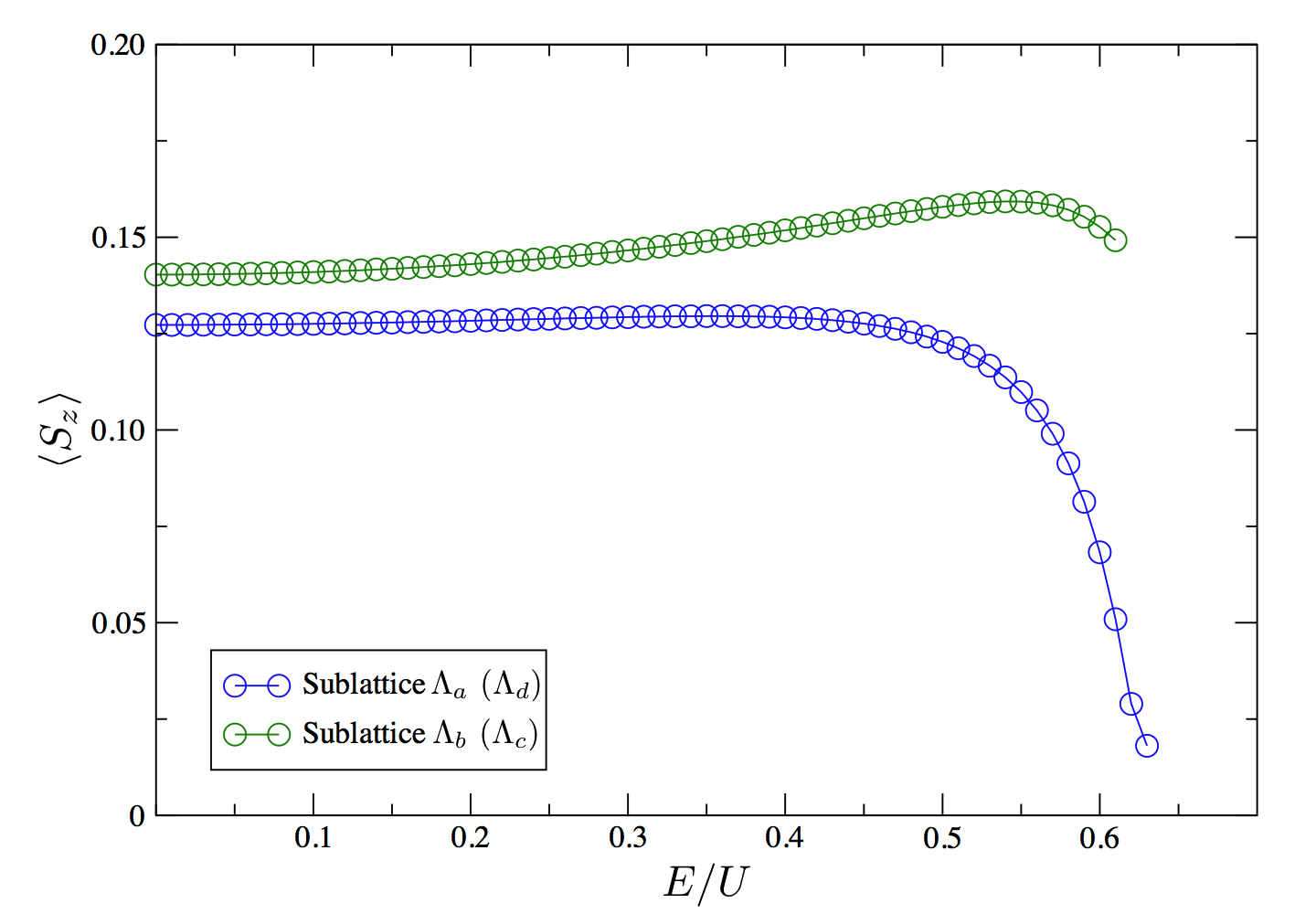}
\caption{(Color online) Magnetization as a function of electric field coupling, $E$. Parameters used are $t_\parallel / U = 0.1$, $t_\perp /U = 0.07$, $V_\parallel /U = 0.4$, and $V_\perp /U = 0.3$. According to Eq.~\ref{eq:couplings1} and \ref{eq:couplings2}, these parameters match the exchange coupling values used in Fig.~\ref{fig:mag-j}.}
\label{fig:mag-e}
\end{figure}

To obtain the staggered magnetization for more realistic states, including the ones in experiments, we need to consider the change of $J_\perp$ and $J_\times$ in a consistent manner. This is done by tuning a single parameter $E$, the coupling of electric field. Using the expressions in Eq.~\ref{eq:couplings1} and \ref{eq:couplings2}, we can find the magnetizations for each sublattice as a function of $E$. As for the previous results, we only show numerical results for 
selected parameters. Fig.~\ref{fig:mag-e} shows the result for the same parameters as in Fig.~\ref{fig:mag-j}, $t_\parallel / U = 0.1$, $t_\perp /U = 0.07$, $V_\parallel /U = 0.4$, and $V_\perp /U = 0.3$. 
We observe the magnetization of $\Lambda_a$ decrease drastically from $E \sim 0.50 U$ and that of $\Lambda_b$ starts to decrease from $E \sim 0.55 U$, although we cannot see a significant decrease in $\Lambda_b$ before the Holstein-Primakoff theory breaks down. However from the two plots in Fig.~\ref{fig:mag-j} where the magnetization of $\Lambda_b$ saturates as increasing $J_\perp$ and vanishes as increasing $J_\times$, we can argue that when both $J_\perp$, $J_\times$ are increasing the magnetization will decrease eventually, and Fig~\ref{fig:mag-e} is showing the onset of the decrease. This result shows explicitly how the N\'{e}el order decreases as the electric field increases.

\section{$J_1$-$J_2$ model}
\label{sec:j1j2}

The fact that the magnetization of $\Lambda_a$ and $\Lambda_d$ decreases faster than that of $\Lambda_b$ and $\Lambda_c$ in the previous section can be taken as evidence that, in the phase transition we are concerned with, it is sufficient to consider sublattices $\Lambda_b$ and $\Lambda_c$ in the effective theory, {\em i.e.\/} the effective single layer honeycomb depicted in Fig.~\ref{fig:blg}. The spins in $\Lambda_a$ and $\Lambda_d$ will form singlets, while $\Lambda_b$ and $\Lambda_c$ still remain in the N\'{e}el phase and remain the important degrees of freedom.

So now we want to directly study an effective model for only the sites in sublattices $\Lambda_b$ and $\Lambda_c$. Following again the strong coupling limit, the resulting theory will also become a spin model. We write the $J_1$-$J_2$ spin model for the effective honeycomb lattice. That is, 
\begin{align}
	H &= J_1 \sum_{i=1}^3 \sum_{\br \in \Lambda_b} \vec{S}^{(b)}_{\br} \cdot  \vec{S}^{(c)}_{\br+\bs_i}  \nonumber \\ 
	&+ J_2 \sum_{i=1}^3 \left[ \sum_{\br \in \Lambda_b}  \vec{S}^{(b)}_{\br} \cdot  \vec{S}^{(b)}_{\br+ \bt_i} 
	+\sum_{\br \in \Lambda_c}  \vec{S}^{(c)}_{\br} \cdot  \vec{S}^{(c)}_{\br+ \bt_i} \right].
	\label{eq:j1j2}
\end{align}
The $J_2$ coupling is $t^4/U^3$ order in the perturbation in Sec.~\ref{sec:model}, and is calculated in Eq.~\ref{eq:couplings1}. However, from the lattice structure in Fig.~\ref{fig:blg}, one can see that $J_1$ is in $t^6/U^5$ order in the same perturbation theory. Calculating perturbation in two extra orders is a straightforward but tedious task, so we seek an alternative way to compute $J_1$. We do this by assuming $J_\perp \gg J_\parallel$, and perform the 
perturbation expansion in $J_\parallel /J_\perp$. Admittedly, because $t_\parallel$ is actually significantly smaller than $t_\perp$ in graphene, this is perturbation
expansion is rather far from the experimental situation; however, the regime $J_\parallel \gg J_\perp$ offers a tractable limit for studying the phase transition using
existing results so seems worthwhile to explore. In the opposite limit of $J_\perp \ll J_\parallel$, qualitatively the magnetization of $\Lambda_a$ and $\Lambda_b$ will be the same although they may be small. Therefore our assumption of $J_\perp \gg J_\parallel$ will be true in regions where $\langle S_z^{(b)}\rangle \gg \langle S_z^{(a)}\rangle$. In Fig.~\ref{fig:mag-e}, this is the case when $E/U > 0.55$. This means that the large $J_\perp$ limit is more valid near the phase transition, and thus suits our purpose of studying the vicinity of the transition point.

The $J_\parallel/J_\perp$ expansion has two contributions to the effective honeycomb lattice in the order of $J_\parallel^2/J_\perp$, one to the $J_1$ term and the other to the $J_2$ term. The contributions from the $J_\parallel/J_\perp$ expansion follows from the effective Hamiltonian method \cite{cohen},
\begin{align}
J_1 &= \frac{J^2_{\parallel}}{J_{\perp}} = \frac{4~t^4_{\parallel}}{t^2_{\perp}} \frac{U-V_{\perp} - \left( \frac{E^2}{U-V_{\perp}} \right)}{\left(U-V_{\parallel} - \left( \frac{V^2_{\perp}}{U-V_{\parallel}} \right) \right)^2} ,\nonumber \\
	J_2 &= -\frac{J^2_{\parallel}}{2 J_{\perp}} =  -\frac{2~t^4_{\parallel}}{t^2_{\perp}} \frac{U-V_{\perp} - \left( \frac{E^2}{U-V_{\perp}} \right)}{\left(U-V_{\parallel} - \left( \frac{V^2_{\perp}}{U-V_{\parallel}} \right) \right)^2}. 
\label{eq:large-jp}
\end{align}
From our assumption that $J_\parallel$ and $J_\perp$ are antiferromagnetic, it follows that the contribution to $J_1$ is antiferromagnetic and $J_2$ is ferromagnetic. For a complete description for the $J_1 - J_2$ model in the effective honeycomb lattice up to the desired order, we need to add the $J_2$ contributions from the $t/U$ expansion and $J_\parallel/J_\perp$ expansion. The final $J_1 - J_2$ model will be Eq.~\ref{eq:j1j2} with exchange couplings of,
\begin{align}
	J_1 =& \frac{4~t^4_{\parallel}}{t^2_{\perp}} \frac{U-V_{\perp} - \left( \frac{E^2}{U-V_{\perp}} \right)}{\left(U-V_{\parallel} - \left( \frac{V^2_{\perp}}{U-V_{\parallel}} \right) \right)^2}, \nonumber \\
	J_2 =& \frac{4~t^4_{\parallel}}{(U-V_{\parallel})^2 - V^2_{\perp}} \left( \frac{2(U-V_{\parallel}) \left( (U-V_{\parallel})^2 + V^2_{\perp}\right)}{\left( (U-V_{\parallel})^2 - V^2_{\perp} \right)^2} -\frac{1}{U} \right) \nonumber \\
	&-\frac{2~t^4_{\parallel}}{t^2_{\perp}} \frac{U-V_{\perp} - \left( \frac{E^2}{U-V_{\perp}} \right)}{\left(U-V_{\parallel} - \left( \frac{V^2_{\perp}}{U-V_{\parallel}} \right) \right)^2}.
	\label{eq:j1j2couplings}
\end{align}

The ground state of the above $J_1$-$J_2$ model can only be solved numerically. However, qualitative behaviors can be studied from the $E$ dependence of $J_1$ and $J_2$. Directly from Eq.~\ref{eq:j1j2couplings}, one can see that $J_1$ decreases and $J_2$ increases as $E$ increases. Since the first term of $J_2$ in Eq.~\ref{eq:j1j2couplings} is positive, we always have a window of $E$ where both $J_1$ and $J_2$ are positive. Inside that window, the ratio of $J_2/J_1$ will increase 
as $E$ increases, until $J_1$ decreases to 0. We know that for $J_2/J_1 \ll 1$ the ground state will be a N\'{e}el state, including when $J_2 <0$ where $J_2$ supports the N\'{e}el state. However a positive $J_2$ starts to frustrate the N\'{e}el phase as $J_2/J_1$ increases. This will eventually destroy the N\'{e}el state at a critical value of $J_2/ J_1$, and a phase transition will occur. 

Numerically, the $J_1$-$J_2$ model in a honeycomb lattice has recently been 
investigated via a variety of methods \cite{clark,lauchli,ganesh,zhu,shengfisher}, and related models have been studied in Refs.~\onlinecite{damle,langsun}.
These studies all find a 
transition out of the N\'eel state to a Kekul\'{e} VBS state (or the closely related plaquette state which has the same pattern on symmetry breaking
on the honeycomb lattice). Refs.~\onlinecite{ganesh,zhu,shengfisher} tune $J_2/J_1$, and find evidence for an apparent 
second order phase transition from N\'{e}el state at small $J_2/J_1$ to VBS state at larger $J_2 / J_1$, where the critical value $J_2 / J_1 \sim 0.22$---$0.26$. 
The studies can be therefore considered as the numerical analysis of our $J_1$-$J_2$ model in the window of $E$ where $J_1,~J_2 >0$. 
Since the critical value of $J_2/J_1$ in the DMRG study can be always reached in our model through a certain value of $E$, we may argue that the same phase transition from N\'{e}el to VBS happens in the bilayer graphene system as well, when tuning the electric field. So the $J_1$-$J_2$ model in the effective honeycomb lattice not only supports the N\'{e}el to VBS phase transition in the bilayer graphene, but also provides indirect evidence that the transition is in the deconfined category. 

\section{Geometric phases}
\label{sec:topo}

Our analysis so far has examined the potential instability of the N\'eel phase to a `quantum disordered' phase which 
preserves spin rotation invariance. General arguments were made in Ref.~\onlinecite{vbsprb} that any such phase in a model with
the symmetry of the honeycomb lattice must have 
VBS order: these arguments relied on Berry phases of `hedgehog' tunneling events in the N\'eel order.
In Ref.~\onlinecite{fu} (see also Ref.~\onlinecite{yao}) 
these arguments were recast in terms of geometric phases associated with skyrmion textures, which led to a coupling in the action
between the temporal derivative of the VBS order and the skyrmion density in the N\'eel order. 
This section will obtain a similar term for the case of the bilayer antiferromagnet. This term will be obtained in a weak coupling model, and we will comment on the relationship to the strong coupling results at the end of the present section.

Since we already know the ground states around the critical point are N\'{e}el and VBS state, we write a weak coupling Hamiltonian and later include interaction effects and the electric field as a N\'{e}el and VBS mean field order parameter. The weak coupling Hamiltonian in a bilayer honeycomb lattice is merely a tight-binding model. Using the parameters and operators defined as in Sec.~\ref{sec:model}, this is, 
\begin{align}
	H_{\textrm w} =& -t_{\parallel} \sum_{\br \in \Lambda_a} \sum_{i=1}^3 c^{(a)\dagger}_{\br}  c^{(b)}_{\br+\bs_i}  
	-t_{\parallel} \sum_{\br \in \Lambda_d} \sum_{i=1}^3 c^{(d)\dagger}_{\br}   c^{(c)}_{\br-\bs_i}   \nonumber \\
	&-t_{\perp} \sum_{\br \in \Lambda_a} c^{(a) \dagger}_{\br} c^{(d)}_{\br}
	-t_2 \sum_{\br \in \Lambda_b} \sum_{i=1}^3 c^{(b)\dagger}_{\br}  c^{(c)}_{\br+\bs_i}
	+\textrm{h.c.}.
	\label{eq:blg_h0}
\end{align}
One extra term is added to Eq.~\ref{eq:hubbard_H}, which is the $t_2$ term describing the direct hopping between sublattice $\Lambda_b$ and sublattice $\Lambda_c$. Although $t_2$ is very small compared to $t_{\parallel}$ and $t_{\perp}$ in realistic systems as we ignored in the previous calculations, we keep the $t_2$ term in the current section to use it as a parameter interpolating between bilayer and monolayer graphene \cite{falko}. 

The band structure of this Hamiltonian consists of four bands where two of them quadratically touches at the two $K$ points which we label them as $K_\pm = \pm (0, \frac{4\pi}{3 \sqrt3})$. At half filling the Fermi level is right at the touching points, and the low energy physics are govern by the $K_\pm$ points of the quadratically touching bands. Also at the $K_\pm$ points, the band gap between the quadratically touching bands and the remaining bands are $t_\perp$. Therefore by considering energies much smaller than $t_\perp$ near the $K_\pm$ points, we write an low energy effective theory, 
\begin{align}
	H_{\textrm w}^{\textrm{eff}} = 
	\sum_{\bf p}
	\Psi^{\dagger} ({\bf p})
	&\left[ 
	\frac{v^2}{t_\perp}\left(  \left( p_x^2 - p_y^2 \right) s_x 
	+  \left( 2 p_x p_y \right) \rho_z s_y \right) \right. \nonumber \\
	&\left. \phantom{\frac{a}{b}more}+ v_2 \left( p_x s_y + p_y \rho_z s_x \right)
	\right]
	\Psi({\bf p}),
	\label{kineticenergy}
\end{align}
where $v= 3t_{\parallel}/2$ and $v_2 = 3 t_2 / 2$. Here, $p_x$ and $p_y$ are the momentum measured from the $K_\pm$ points, $\rho$ and $s$ are the Pauli matrices in valley and layer space, respectively. Only sublattice $\Lambda_b$ and $\Lambda_c$ remain in the effective theory, and $\Psi ({\bf p})$ is a four component spinor with each component from two sublattices and two $K_\pm$ points. $\Lambda_b$ and $\Lambda_c$ also forms a honeycomb lattice and we again see that the effective low energy theory of a bilayer honeycomb lattice lives in a single honeycomb lattice. 

Now we impose the system is in N\'{e}el phase. In the ordered state, we may choose the N\'{e}el order parameter to be in $z$-direction, and we can simply add $H_{N_{z}} = m \sigma_z s_z$ to the effective Hamiltonian, where $\sigma$ are the spin Pauli matrices. The N\'{e}el order opens up a gap of size $2m$ at the $K_\pm$ points. $H_0 = H_{\textrm w}^{\textrm{eff}} + H_{N_{z}}$ is the final effective Hamiltonian for the system in the N\'{e}el phase and will serve as the unperturbed Hamiltonian. 

As the system approaches the critical point, N\'{e}el order and VBS order fluctuation becomes larger. Therefore, as in Ref.~\onlinecite{fu}, both fluctuations should be taken into account for a proper study of the system near the critical point. We treat these two as a perturbation. Let us write the fluctuating N\'{e}el order parameter as ${\vec m} = m(n_x , \, n_y ,\, 1)$ and the complex VBS order parameter as $V = V_x + \i V_y$. The Hamiltonian of $n_x , \, n_y$ is $H_{N_{xy}}=m s_z\left(n_x \sigma_x + n_y \sigma_y \right)$. Recalling that Kekul\'{e} type of bond order can be written as a modulation on the tight binding hopping parameter, \cite{hou-prl, hou-prb}
\begin{align}
	H_V &=  -\sum_{\br \in \Lambda_b} \sum^{3}_{i =1} \delta t_{{\br},i}~ c^{(b) \dagger}_{\br} c^{(c)}_{\br + \bs_i} + \textrm{h.c.}, \\
	\delta t_{{\br},i} &= V~ \e^{\i {\bf K}_+ \cdot {\bs}_i}  \e^{\i ( {\bf K}_+ - {\bf K}_-)  \cdot {\br}}/3 + \textrm{c.c.}, \nonumber
\end{align}
we find $H_V = -s_x \left( V_x \rho_x - V_y \rho_y \right)$ as the Hamiltonian for the VBS order parameter. So the perturbation $H_1$ is $H_1 = H_{N_{xy}}+H_V$ and now we can write the full Hamiltonian,
\begin{align}
	H &= H_0 + H_1 \nonumber \\
	&= \left(
	\frac{v^2}{t_\perp}\left(  \left( p_x^2 - p_y^2 \right) s_x 
	+  \left( 2 p_x p_y \right) \rho_z s_y \right)
	+ v_2 \left( p_x s_y + p_y \rho_z s_x \right) \right. \nonumber \\
	&\left. \phantom{\frac{v^2}{t_\perp}} +  m \sigma_z s_z \right) 
	+ \left(
	m \left(n_x \sigma_x + n_y \sigma_y \right) s_z
	- \left( V_x \rho_x - V_y \rho_y \right) s_x
	\right).
	\label{eq:full_H}
\end{align}
Note that the terms proportional to the antiferromagnetic order, $m$, anti-commute with all the terms in 
Eq.~(\ref{kineticenergy}), indicating they will open up a gap in the electronic spectrum. On the other hand, the terms proportional to the VBS
order anti-commute only with the $v_2$ term, but not with the $v^2 / t_\perp$ term, indicating that VBS order alone does not open a gap
in the purely quadratic-band-touching spectrum.

Writing in a specific basis, $\Psi^{\dagger} ({\bp}) = (c^{(b) \dagger}_{{\bp}+}, \, c^{(b) \dagger}_{{\bp}-}, \, c^{(c) \dagger}_{{\bp}+}, \, c^{(c) \dagger}_{{\bp}-})$, where $\pm$ corresponds to the $K_\pm$ points the momentum is measured from, 
\begin{align}
	&H =  \nn 
	&\left( \begin{array}{cccc} 
	{\vec m }\cdot {\vec \sigma}& 0 & -\frac{v^2}{t_\perp} \pi^2 + v_2 \pi^\dagger& -V_x - \i V_y \\ 
	0 & {\vec m }\cdot {\vec \sigma}& -V_x + \i V_y & -\frac{v^2}{t_\perp} \pi^{\dagger\, 2} - v_2 \pi \\ 
	 -\frac{v^2}{t_\perp} \pi^{\dagger\,2} + v_2 \pi & -V_x - \i V_y & -{\vec m }\cdot {\vec \sigma}& 0  \\ 
	-V_x + \i V_y & -\frac{v^2}{t_\perp} \pi^2 - v_2 \pi^\dagger & 0 & -{\vec m }\cdot {\vec \sigma}\\ 
	\end{array} \right).		
	\label{eq:hblg_basis}
\end{align}
Here, $\pi = \i p_x + p_y$ is defined for notational convenience. Now it is more apparent that $v_2 = 0$ gives the Hamiltonian for bilayer graphene and $v = 0$ gives that of the monolayer graphene with opposite chirality. Therefore, we may tune $v_2 / v$ to interpolate between monolayer and bilayer graphene. 

Note that the electric field $E$ in Eq.~\ref{eq:hubbard_H} is not included in this final form of the Hamiltonian. However, it is encoded in the order parameters as we have seen in the previous sections how electric field tunes the N\'{e}el to VBS transition. The electric field has other effects as well, as changing the energy gap of the VBS phase for example, but this will have only quantitative effects in the calculation. The result of the calculation with explicit electric field will be presented in Appendix \ref{sec:apx_A}.

From Eq.~\ref{eq:full_H}, we integrate out the fermions to get an effective theory for the fluctuating order parameters. The coupling between the N\'{e}el and VBS order parameters appears at fourth order of one-loop expansion. For notational simplicity, we follow Ref.~\onlinecite{fu} and combine the four real order parameter to a multicomponent bosonic field, $\mathcal{A}^{\mu} (x,y, \tau) = (V_x /m ,\, V_y /m ,\,  n_x ,\,  n_y)$. $\mu = 0, 1, 2, 3$ labels the different fields in $\mathcal{A}^{\mu}$ and are not Lorentz indices. In momentum space, the four point coupling between the bosonic fields are, 
\begin{align}
	S_1 = \sum_{\mu, \nu, \lambda, \delta} \int \prod_{i=1}^{3} d {p}_i K^{\mu \nu \lambda; \delta}_{{p}_1 {p}_2 {p}_3}
		\mathcal{A}^{\mu} ({p}_1 ) \mathcal{A}^{\nu}({p}_2) \mathcal{A}^{\lambda}({p}_3) \nn
		\phantom{space} \times \mathcal{A}^{\delta}(-{p}_1 -{p}_2 -{p}_3  ).
	\label{eq:s1}
\end{align}
Among this bosonic coupling, we are most interested in the topological term,
\begin{align}
	S_{\textrm{top}}	 &= \i \int dx dy d\tau  \left(K j_{\tau}^N j_{\tau}^V + K' j_{x}^N j_{x}^V + K' j_{y}^N j_{y}^V \right),
	\label{eq:topo_action}
\end{align}
which is the coupling term between the skyrmion current, $j_{\alpha}^N$, and VBS current, $j_{\beta}^V$.
\begin{align}
	j_{\alpha}^N &\equiv \epsilon_{\alpha \beta \gamma} \epsilon_{a b c} n^a \partial_\beta n^b \partial_\gamma n^c \nonumber \\
	j_{\beta}^V &\equiv V_x \partial_\beta V_y - V_y \partial_\beta V_x
\end{align}
This topological term is of interest to us because it provides an argument that the system is in VBS phase in the disordered side; as mentioned in the beginning of this section, this is analogous to arguments in Refs.~\onlinecite{vbsprb} and \onlinecite{fu}.

\begin{figure}[t]
	\vspace*{2mm}
	\includegraphics*[width=2.7in,angle=0]{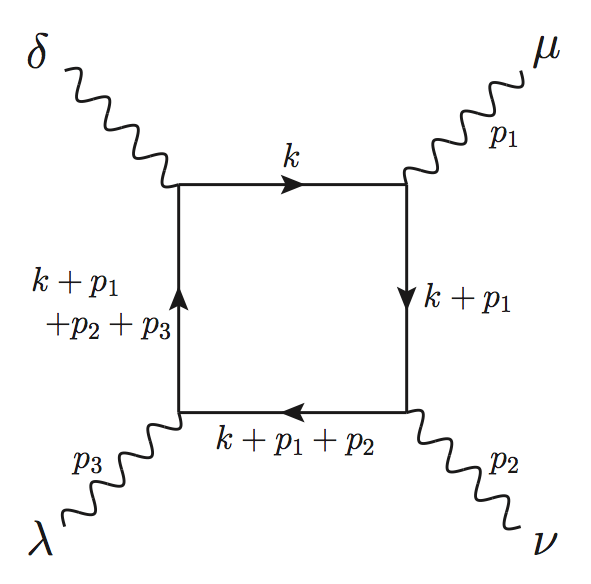}
	\caption{The box diagram needed for the calculation of $K$ and $K'$. The $p_1^\alpha p_2^\beta p_3^\gamma$ coefficient of this box diagram gives $K^{\mu \nu \lambda; \delta}_{\alpha \beta \gamma}$ which $K$, $K'$ consist of. The exact relations between $K$ and $K^{\mu \nu \lambda; \delta}_{\alpha \beta \gamma}$ is given in Eq.~\ref{eq:k}. Note that every momentum dependence comes from the propagator.}
	\vspace*{2mm}
	\label{fig:diagrams}
\end{figure}

As explained in detail in Ref.~\onlinecite{fu}, we can extract the couplings $K$ and $K'$ from $K^{\mu \nu \lambda; \delta}_{{p}_1 {p}_2 {p}_3}$ in Eq.~\ref{eq:s1}. The final expression for $K$ is as follow, 
\begin{align}
	8  K &= K^{2 3 4; 1}_{\tau xy} + K^{2 4 3; 1}_{\tau yx}+K^{3 4 2; 1}_{xy \tau}  + K^{3 2 4; 1}_{x \tau y} + K^{4 2 3; 1}_{y \tau x} + K^{4 3 2; 1}_{y x \tau} \nonumber \\
	&\quad -\left( K^{2 4 3; 1}_{\tau xy} + K^{2 3 4; 1}_{\tau yx}+K^{4 3 2; 1}_{xy \tau}  + K^{4 2 3; 1}_{x \tau y} + K^{3 2 4; 1}_{y \tau x} + K^{3 4 2; 1}_{y x \tau} \right) \nonumber \\
	&\quad -\left( K^{1 3 4; 2}_{\tau xy} + K^{1 4 3; 2}_{\tau yx}+K^{3 4 1; 2}_{xy \tau}  + K^{3 1 4; 2}_{x \tau y} + K^{4 1 3; 2}_{y \tau x} + K^{4 3 1; 2}_{y x \tau} \right) \nonumber \\
	&\quad + K^{1 4 3; 2}_{\tau xy} + K^{1 3 4; 2}_{\tau yx}+K^{4 3 1; 2}_{xy \tau}  + K^{4 1 3; 2}_{x \tau y} + K^{3 1 4; 2}_{y \tau x} + K^{3 4 1; 2}_{y x \tau} ,
	\label{eq:k}
\end{align}
where $K'$ can also be written in a similar way. Here, $K^{\mu \nu \lambda; \delta}_{\alpha \beta \gamma}$ are defined as the coefficient of the term linear in $p_1 p_2 p_3$. 
\begin{align}
	K^{\mu \nu \lambda; \delta}_{{p}_1 {p}_2 {p}_3} = \cdots + K^{\mu \nu \lambda; \delta}_{\alpha \beta \gamma} p_1^\alpha p_2^\beta p_3^\gamma+ \cdots
\end{align}
The lowest order contribution to $S_1$ arises from the one-loop expansion, when we integrate out the fermion loop. Therefore, the calculation of $K^{\mu \nu \lambda; \delta}_{\alpha \beta \gamma}$ eventually boils down to calculating box diagrams, as in Fig.~\ref{fig:diagrams}. Note that in Eq.~\ref{eq:full_H}, the vertex functions between bosonic fields and fermions have no momentum dependence and therefore the $p_1^\alpha$, $p_2^\beta$, $p_3^\gamma$ dependence comes from the propagator. 

After evaluating a number of diagrams and substituting $K^{\mu \nu \lambda; \delta}_{\alpha \beta \gamma}$'s to Eq.~\ref{eq:k} and its $K'$ analog, we obtain the topological couplings $K$ and $K'$ of the system. First, we consider the bilayer limit of $v_2/v = 0$. The integral expressions for $K$ and $K'$ are, 
\begin{widetext}
\begin{align}
	K & =\frac{1}{8\pi^3} \int d k_0 d k_x d k_y  \frac{8 m^5 t_{\perp}^6 v^4 \left( k_x^2 + k_y^2 \right)}{\left( (k_0^2 + m^2)t_{\perp}^2 + (k_x^2 + k_y^2)^2 v^4\right)^4}, \nonumber \\ 
	K' &= \frac{1}{8\pi^3}\int d k_0 d k_x d k_y \frac{4 m^5 t_{\perp}^6 v^4 \left( 3 (k_x^2 - k_y^2) (k_0^2 + m^2)t_{\perp}^2  -  (k_x^2 + k_y^2)^2 (13 k_x^2 + 3k_y^2 ) v^4 \right)  }{\left( (k_0^2 + m^2)t_{\perp}^2 + (k_x^2 + k_y^2)^2 v^4 \right)^5} \nonumber \\
	&= \frac{1}{8\pi^3}\int d k_0 d k_x d k_y \frac{4 m^5 t_{\perp}^6 v^4 \left( 3 (-k_x^2 + k_y^2) (k_0^2 + m^2)t_{\perp}^2  -  (k_x^2 + k_y^2)^2 (3 k_x^2 + 13k_y^2 ) v^4 \right)  }{\left( (k_0^2 + m^2)t_{\perp}^2 + (k_x^2 + k_y^2)^2 v^4 \right)^5}.
\end{align}
\end{widetext}
The first expression for $K'$ is the coupling of $j_x^N j_x^V$ and the second is of $j_y^N j_y^V$. They map to each other by the transformation $k_x \leftrightarrow k_y$ and gives the same value when integrated on a region which has $k_x \leftrightarrow k_y$ symmetry as well. We can obtain the $K$ and $K'$ of the effective theory by integrating $k_x$ and $k_y$ in whole space \cite{abanov}. In zero temperature, performing the $k_0$, $k_x$, and $k_y$ integral gives $K = -1 /16 \pi $ and $K' = 1 /16 \pi$. This is a quantized value which does not depend on microscopic parameters $m$, $t_{\perp}$, or $v$.

\begin{figure}[b]
	\includegraphics*[width=3.1in,angle=0]{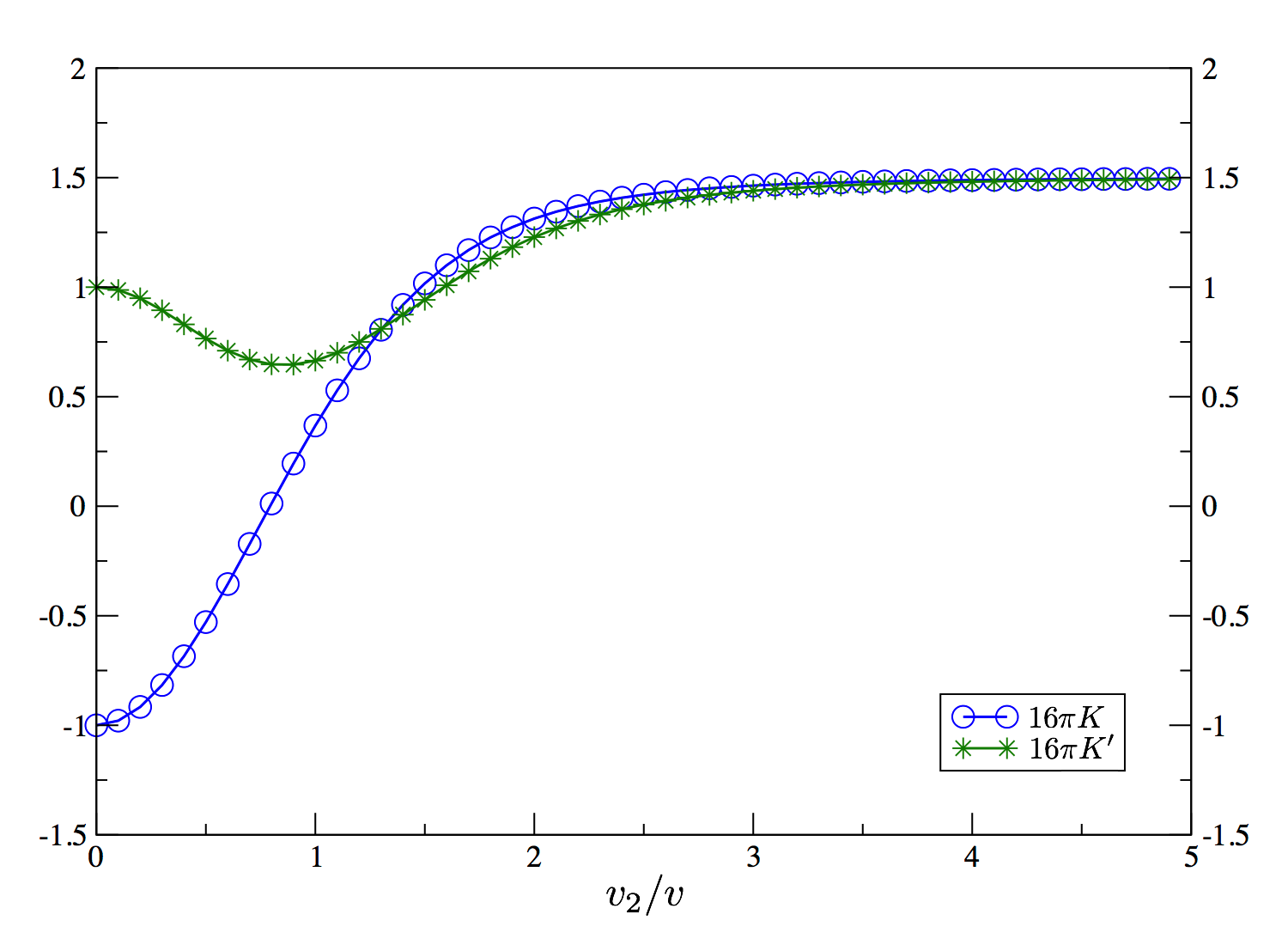}
	\caption{(Color online) The topological couplings $K$ and $K'$ as a function of $v_2 / v$. We assume $ t_{\parallel} = t _{\perp}= m$, and the $y$-axis is in the unit of $1/ 16\pi$. In the bilayer limit of $v_2 / v = 0$, the couplings are $K=- 1/ 16 \pi$ and $K' = 1/ 16\pi$. Both $K$ and $K'$ approaches $3/ 32\pi$ in the monolayer limit of $v_2 / v  \gg 0$.}
	\label{fig:k_and_kp}
\end{figure}

Next we consider the monolayer limit of $v_2/v \gg 1$. This can be integrated analytically and gives $K = K' = 3/32\pi$. They are quantized as well as in the bilayer limit, and the values are consistent with the result from Ref.~\onlinecite{senthil3}. We also compute $K$ and $K'$ as a function of $v_2/v$ and observe how it changes in the intermediate regime of monolayer and bilayer. The result in Fig.~\ref{fig:k_and_kp} shows the aforementioned limiting values of $K = K' = 3/32\pi$ for monolayer and $K = -K' = -1/16\pi$ for bilayer, and a continuous interpolation in-between. $K$ and $K'$ of the intermediate regime does not remain quantized, and thus depends on the parameters of the theory. Through a number of numerical calculations, we observe that the intermediate values at a given $v_2/v$ depends on a single parameter $m/t_{\perp}$. Starting from the bilayer ($v_2/v = 0$), the convergence to the single layer limit occurs more rapidly for smaller values of $m/t_{\perp}$. 

The presence of these non-zero terms supports the proposal that the general structure of the coupling between the N\'eel and VBS orders
is the same as that in the single layer honeycomb lattice. But the values of the geometric phases differ in the weak coupling theory,
although we expect them to coincide in the strong coupling theory (by the arguments of Ref.~\onlinecite{vbsprb}).
This difference suggests that the weak coupling analysis points to a first-order N\'eel-VBS transition, while deconfined
criticality is preferred at strong coupling.

\section{Zero mode in VBS vortex}
\label{sec:vortex}

Another approach to the theory of deconfined criticality is via the structure of vortices in the VBS order parameter.
Levin and Senthil \cite{levin} presented general arguments that each such vortex must carry spin $S=1/2$.
In some case, this fact is already apparent by the presence of zero modes in a weak coupling theory of the VBS state:
this is the case in monolayer graphene \cite{hou-prl, hou-prb}. We present a corresponding computation for the bilayer case,
and do not find such zero modes. We view this as a feature of the weak coupling approach, rather than 
the inapplicability of the
general strong coupling arguments of Ref.~\onlinecite{vbsprb}.

We start in the deep VBS phase and may ignore the N\'{e}el order for now. Therefore we adapt the Hamiltonian Eq.~\ref{eq:full_H} with $m=0$. In this section we return to $t_2 =0$, and set the Fermi velocity $v = 3t_{\parallel}/2$ equal to one. VBS order parameter, $V (\br) = V_x (\br)+ \i V_y( \br)$, now has a nonzero expectation value, and let us allow it to fluctuate over space. We will assume the fluctuation is in a much longer length scale than the lattice constant, which is set to 1, so we can treat the order parameter as a constant during Fourier transform. Moreover, we include the electric field explicitly as this will open up an energy gap of the system. The Hamiltonian in the VBS phase is, 
\begin{align}
	H =& 
	\frac{1}{t_\perp} \left( p_x^2 - p_y^2 \right) s_x 
	+ \frac{1}{t_\perp} \left( 2 p_x p_y \right) \rho_z s_y \nonumber \\
	&- \left( V_x (\br) \rho_x - V_y (\br) \rho_y \right) s_x 
	- E s_z,
	\label{eq:H_vbs}
\end{align}
where $\rho$ and $s$ are the Pauli matrices in valley and layer space, respectively. Note that in the bilayer graphene, the Kekul\'{e} VBS order does not open up a gap but create nodal lines which form a circle in the Brillouin zone. The nodal line is protected by the layer symmetry of the system. Discussing zero modes in a gapless system is meaningless, however, the transition is at a nonzero electric field. Electric field couples differently with the density of the electrons in different layers and breaks the layer symmetry. This opens up a gap at the nodal line and it is now legitimate to discuss zero modes of the system. 

As in Sec.~\ref{sec:topo}, choosing a specific basis, $\Psi^{\dagger} ({\bp}) = (c^{(b) \dagger}_{{\bp}+}, \, c^{(b) \dagger}_{{\bp}-}, \, c^{(c) \dagger}_{{\bp}+}, \, c^{(c) \dagger}_{{\bp}-})$, we can represent the Hamiltonian as a four by four matrix. 
\begin{align}
	H = \left( \begin{array}{cccc} 
	\vspace{1mm}
	-E ~&~ 0 ~&~ - \frac{1}{t_\perp} \pi^2 ~&~ - V ({\br}) \\ 
	\vspace{1mm}
	0 ~&~ -E ~&~ - \overline{V} ({\br}) ~&~ -\frac{1}{t_\perp} \pi^{\dagger\, 2} \\ 
	\vspace{1mm}
	-\frac{1}{t_\perp} \pi^{\dagger \,2} ~&~ - V ({\br}) ~&~ E ~&~ 0  \\ 
	- \overline{V} ({\br}) ~&~ -\frac{1}{t_\perp} \pi^2 ~&~ 0 ~&~ E \\ 
	\end{array} \right)	
\end{align}
Changing to the real basis, $\Psi^{\dagger}({\br}) = (u_b^{\dagger} ({\br}), v_b^{\dagger} ({\br}) , u_c^{\dagger} ({\br}),  v_c^{\dagger} ({\br}) )$, where for example, $u_b ({\br}) = \frac{1}{\sqrt N} \sum_{{\bp}} \e^{\i {\bp} \cdot {\br}} b_{{\bp}+}$ and $v_b ({\br}) = \frac{1}{\sqrt N} \sum_{{\bp}} \e^{\i {\bp} \cdot {\br}} b_{{\bp}-}$, 
\begin{align}
	H = -\frac{1}{t_\perp} \left( \begin{array}{cccc} 
	\vspace{1mm}
	E ~&~ 0 ~&~ 4 \partial_z^2  ~&~ V({\br}) \\ 
	\vspace{1mm}
	0 ~&~ E ~&~ \overline{V} ({\br}) ~&~ 4 \partial_{\bar{z}}^2 \\ 
	\vspace{1mm}
	4 \partial_{\bar{z}}^2 ~&~ V ({\br}) ~&~ -E ~&~ 0  \\ 
	\overline{V} ({\br}) ~&~ 4 \partial_z^2  ~&~ 0 ~&~ -E \\ 
	\end{array} \right).
\end{align}
Here, we included $t_\perp$ into the definition of $V ({\br})$ and $E$ for notational convenience and used complex coordinate $z = x+ \i y$ for $2 \partial_z = \e^{-\i \theta} (\partial_r - \frac{\i}{r} \partial_\theta)$. 

Now we assume the VBS order parameter contains a vortex, $V(\br) = V_0 (r) \e^{\i \theta}$. Ref.~\onlinecite{jackiw} provides an analytical method of obtaining the zero modes when the Fermions are Dirac-like. They count the number of zero modes by matching the two asymptotic behaviors of the solutions of $H\, \Psi({\br}) = 0$. The quadratic dispersion of Fermions can be easily implemented into this scheme, however, including the electric field ruins the argument and we cannot follow the same step. Alternatively, we have solved the problem numerically. We consider a bilayer honeycomb lattice with $3600$ sites with the corresponding lattice Hamiltonian of Eq.~\ref{eq:H_vbs}, and introduce a vortex at the center of the lattice. Open boundary condition is imposed to deal with the vortex without including any Dirac strings. Introducing a vortex and anti-vortex pair will also resolve the issue, but it will also effectively decrease the system size. With the open boundary condition, we turn on a small potential at the boundary to eliminate zero energy states arising from boundary effects. We then numerically diagonalize the system and search for zero energy eigenvalues. We also check the eigenfunctions of the states while moving the vortex center around to confirm whether the wavefunctions are actually localized at the vortex. The result clearly showed no zero modes in the presence of a vortex.  As we noted earlier, we believe this result is a feature 
of the weak coupling method, and that the needed zero mode will appear in the strong coupling limit as argued in Ref.~\onlinecite{levin}.

We also mention previous reports about zero modes in vortices of bilayer graphene \cite{moon, herbut}. 
In these works, the authors claim there are two zero modes for a single vortex of valley ferromagnet order. Note that this order breaks time reversal symmetry and is different from the Kekul\'{e} VBS phase we have considered above. In the notation of Eq.~\ref{eq:H_vbs}, the valley ferromagnet order will be written as $(V_x \rho_y + V_y \rho_x )s_y$, which anti-commutes with the kinetic energy terms. Ref.~\onlinecite{herbut} concentrates in regions near the vortex and uses the method of Ref.~\onlinecite{jackiw} in momentum space. However, this is potentially dangerous because, as mentioned before, 
the number of zero modes are determined by the matching of the asymptotic behavior of near-vortex and far-vortex regions. 
Indeed, without the matching procedure, and only looking in the near-vortex region one can find an infinite number
of zero modes. However by numerically diagonalizing the lattice Hamiltonian as above, we indeed find two zero modes for their system 
with valley ferromagnet order.

\section{Conclusions}
\label{sec:conc}

Our paper has examined the strong coupling limit of an extended Hubbard model appropriate for undoped bilayer graphene.
The results of our analyses are that the application of a transverse electric field does indeed destabilize the N\'eel insulator,
and that resulting `quantum disordered' state is likely to have VBS order which breaks the space group symmetry of the lattice.
These results are in accord with the weak coupling analysis \cite{khariprl}, and
our strong coupling arguments indicate that the N\'eel-VBS quantum phase transition
in bilayer graphene can be in the deconfined universality class \cite{senthil1,senthil2}.

On the experimental side, there is now good evidence for the N\'eel state in bilayer graphene \cite{maher,basel}, and
also for a quantum transition out of this state upon application of a transverse electric field \cite{weitz,freitag,macdonaldexp}.
It would be of great interest to devise experiments to measure the translational symmetry breaking associated with the VBS order.
The transition to the VBS state should exhibit quantum-critical scaling, and this may be detected by a careful study
of the temperature dependence of the influence of the transverse electric field on conductance across the transition, and looking for a
``quantum-critical fan'' \cite{keimer} in the electric-field/temperature plane.

For experimental applications, the fundamental new idea that a `deconfined-critical' perspective brings is that the transition
out of the N\'eel state occurs as a consequence of condensation of low energy skyrmions in the N\'eel order;\cite{senthil1,senthil2} 
so such low energy skyrmions should be present in bilayer graphene near the transition. In the presence of ferromagnetic order,  a crucial feature of skyrmions in the
quantum Hall regime is that they carry electric charge.\cite{kanelee,sondhi} In the present bilayer case, each layer has intra-layer ferromagnetism in the 
N\'eel state, with opposite orientation in the two layers, 
and so we can expect that the layers carry opposite charges in the presence of a skyrmion. With the application of an electric field, the layer-exchange symmetry
is broken, and then a skyrmion current will carry a net electrical current. It is notable that the experiments show enhanced conductivity in the region of the transition, \cite{weitz} and this could be explained by the presence of low-energy skyrmions in the deconfined-critical theory. In our present strong-coupling analysis, the
orbital magnetic field effects have been suppressed, and so we have not accounted for the electrical nature of the skyrmions: an extension of our
analysis to include the physics of Landau levels is required, and is being undertaken. On the experimental side, the opposite layers charges carried
by the skyrmion could be studied by driving currents in opposite directions in the two layers. 
Also, optical experiments 
can detect the spin-chirality fluctuations \cite{nagaosa} linked to the collective gauge excitations of deconfined criticality; 
however,
it will be necessary for the light to couple selectively to one layer ({\em i.e.\/} one sublattice of the antiferromagnet).

\begin{acknowledgements}
We thank D.~Abanin, G.-Y. Cho, D. Chowdhury, L. Fu, M.~Kharitonov, Y. Huh, L.-Y. Hung, E. G. Moon, M. Punk, B.~Roy, E.~Shimshoni, and A.~Yacoby for useful 
discussions. This research
was supported by the NSF under Grant DMR-1103860, the Templeton foundation, and MURI grant W911NF-14-1-0003 from ARO.
Research at Perimeter Institute is supported by the
Government of Canada through Industry Canada and by the Province of
Ontario through the Ministry of Research and Innovation.
J. L. is also supported by the STX Foundation.
\end{acknowledgements}

\begin{appendix}

\section{Geometric phases in electric field}
\label{sec:apx_A}

\begin{figure}
\centering
\hspace{0.17in}
\includegraphics[width=3.in]{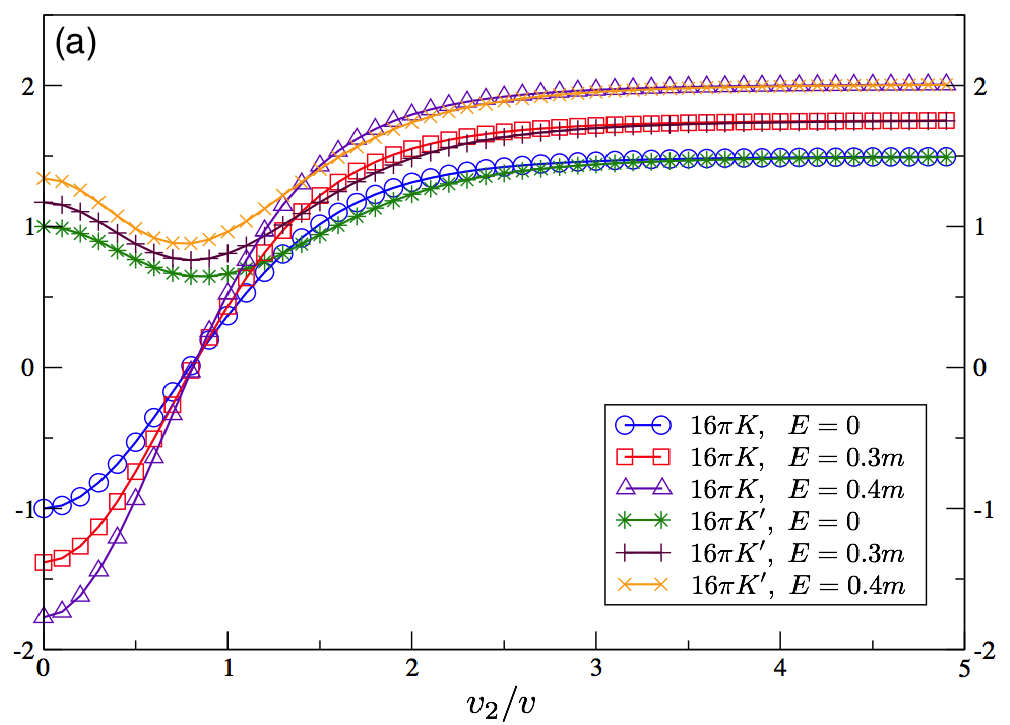}
\includegraphics[width=3.05in]{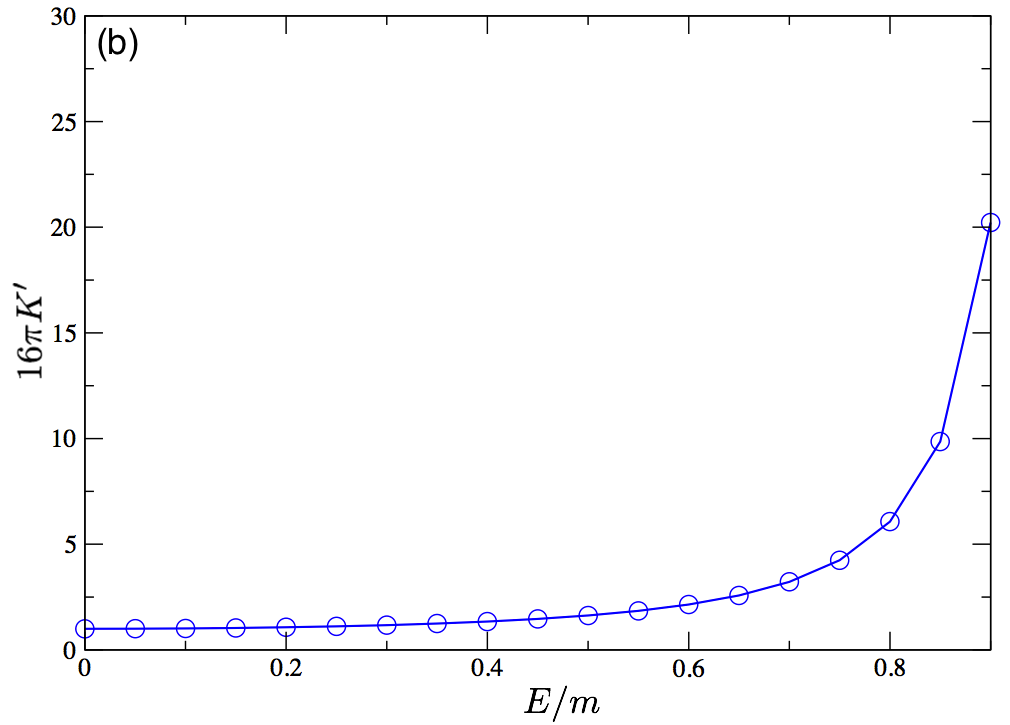}
\caption{(Color online) (a) A plot of $K$ and $K'$ with different values of electric field, as a function of $v_2/v$. We assume $t_{\parallel}= t_{\perp}=m$ and thus the graph with $E=0$ is identical to Fig.~\ref{fig:k_and_kp}. With the nonzero electric field values, the graph shows similar behavior but differences in the exact values. (b) The value of $K'$ in the bilayer limit ($v_2/v= 0$) as we increase the electric field. The value monotonically increases as we increase the electric field. The gap closes at $E=m$ and $K'$, $K$ diverges at this point. }
\label{fig:with_e}
\end{figure}

We revisit the geometric phase calculation in Sec.~\ref{sec:topo} including the electric field to the Hamiltonian. As in Eq.~\ref{eq:H_vbs}, electric field couples to the layer space as an extra $-E s_z$ term. The modified Hamiltonian is, 
\begin{align}
	H &= \left(
	\frac{v^2}{t_\perp}\left(  \left( p_x^2 - p_y^2 \right) s_x 
	+  \left( 2 p_x p_y \right) \rho_z s_y \right)
	+ v_2 \left( p_x s_y + p_y \rho_z s_x \right) \right.  \nonumber \\ 
	&\left. \phantom{\frac{}{}} +  m \sigma_z s_z \right)
	+ \left(
	m \left(n_x \sigma_x + n_y \sigma_y \right) s_z
	- \left( V_x \rho_x - V_y \rho_y \right) s_x
	\right)
	- E s_z.
\end{align}
The details of the calculation are the same as in Sec.~\ref{sec:topo}, where the only difference comes from the new Hamiltonian. 
The couplings $K$ and $K'$ as a function of $v_2/v$ in the presence of electric fields $E=0.3m$ and $E=0.4m$ are shown in Fig.~\ref{fig:with_e}a, together with the $E=0$ case already in Fig.~\ref{fig:k_and_kp}.

$K$ and $K'$ in electric field both shows qualitatively similar behavior to the zero electric field situation. However, one should notice the quantitative values are different not only the intermediate region, but also at the $v_2/v =0$ and $v_2/v \gg0$ limits, where we found 
that the values at $E=0$ were quantized. The reason for this deviation is that the coupling matrices of the N\'{e}el order parameter ($\sigma_i s_z$) and of the electric field ($s_z$) commute, and as a result, the N\'{e}el state and the electric field induce gapped states that 
can mix with each other. Although they have similar dispersion in the weak coupling theory, the two states have very different topological features. For example, the gapped state by electric field does not have topological defects as skyrmions, whose geometric phase leads to nonzero $K$ and $K'$ coupling terms. Explicit calculation also confirms $K=K'=0$ when there is only electric field and no N\'{e}el order parameter in the theory. 

We also calculate $K$ and $K'$ in the bilayer ($v_2/v = 0$) and monolayer ($v_2/v \gg 1$) limit for various values of $E$. In Fig.~\ref{fig:with_e}b, we see that $K'$ of the bilayer limit increases as electric field increase up to $E/m=1$. $E=m$ is the fine tuned value of $E$ where the energy gap vanishes. Both $K$ and $K'$ diverges at this gapless point. Also,
the ratios of $K$ and $K'$ can be written in a simple formula. Let us write $K(v_2/v)$ and $K'(v_2/v)$ as a function of $v_2/v$ for notational convenience. First, the $K(0)$ and $K'(0)$ are related as, 
\begin{align}
	\frac{K(0)}{K'(0)} = - 1 - 2 \left( \frac{E}{m} \right)^2 ,
\end{align}
when $E<m$. This gives the correct limiting value for $E=0$ case, where $K(0) = - K'(0) = -1/16\pi$. Also, as one can check roughly in Fig.~\ref{fig:with_e}a, $K(\infty) = K'(\infty)  = 1.5 K'(0)$ strictly holds as in the $E=0$ limit.

\end{appendix}

\end{document}